\documentclass[twocolumn, twocolappendix, trackchanges]{aastex63}

\usepackage{amsmath}
\usepackage{booktabs}
\usepackage{longtable}
\usepackage{rotating}
\usepackage{graphicx}
\usepackage{hyperref}
\usepackage{bm}

\newcommand{\nc}[2]{\newcommand{#1}{\ensuremath{#2}\xspace}}

\usepackage{xstring} 

\newcommand{\val}[1]{%
  \IfEqCase{#1}{%
{nstars-legacy}{719}
{nstars-new}{135}
}[\PackageError{tree}{Undefined option to tree: #1}{}]%
}%

\usepackage{xspace}

\nc{\specmatch}{ \mathbf{SpecMatch} } 
\nc{\emcee}{ \mathbf{emcee} }
\nc{\radvel}{ \mathbf{RadVel} }
\nc{\rvsearch}{ \mathbf{RVSearch} }
\nc{\thejoker}{ \mathbf{TheJoker} }

\nc{\Kp}{ \textit{Kp} }
\nc{\dAICc}{\Delta\mathrm{AICc}}
\nc{\dBIC}{\Delta\mathrm{BIC}}

\nc{\rsun}{R_{\odot}}
\nc{\msun}{M_{\odot}}
\nc{\rearth}{R_{\oplus}}
\nc{\mearth}{M_{\oplus}}
\nc{\fearth}{F_{\oplus}}
\nc{\mjup}{M_{\textrm{J}}}
\nc{\Rp}{R_P}
\nc{\Mp}{M_P}
\nc{\msini}{M \sin i}
\nc{\Rstar}{R_\star} 
\nc{\Mstar}{M_\star}

\nc{\teff}{T_{\rm eff}}
\nc{\logg}{\log{g}}
\nc{\feh}{[\mbox{Fe}/\mbox{H}]}
\nc{\vsini}{V \sin i}

\nc{\kms}{\text{km\,s}^{-1}}
\nc{\ms}{\text{m\,s}^{-1}}
\nc{\msy}{\text{m\,s}^{-1}\text{yr}^{-1}}
\nc{\gmc}{\text{g\,cm}^{-3}}

\received{August 23rd, 2021}
\revised{April 11th, 2022}
\accepted{May 20th, 2022}
\submitjournal{AAS Journals}

\shortauthors{Rosenthal et al.}
\shorttitle{California Legacy Survey III: On The Shoulders of (Some) Giants}

\begin{document}
\pagenumbering{arabic}

\title{THE CALIFORNIA LEGACY SURVEY III. ON THE SHOULDERS OF (SOME) GIANTS: THE RELATIONSHIP BETWEEN INNER SMALL PLANETS AND OUTER MASSIVE PLANETS}

\correspondingauthor{Lee J.\ Rosenthal}
\email{lrosenth@caltech.edu}

\author[0000-0001-8391-5182]{Lee J.\ Rosenthal}
\affiliation{Cahill Center for Astronomy $\&$ Astrophysics, California Institute of Technology, Pasadena, CA 91125, USA}

\author[0000-0002-5375-4725]{Heather A.\ Knutson}
\affiliation{Division of Geological and Planetary Sciences, California Institute of Technology, Pasadena, CA 91125, USA}

\author[0000-0003-1728-8269]{Yayaati Chachan}
\affiliation{Division of Geological and Planetary Sciences, California Institute of Technology, Pasadena, CA 91125, USA}

\author[0000-0002-8958-0683]{Fei Dai}
\affiliation{Division of Geological and Planetary Sciences, California Institute of Technology, Pasadena, CA 91125, USA}

\author[0000-0001-8638-0320]{Andrew W.\ Howard}
\affiliation{Cahill Center for Astronomy $\&$ Astrophysics, California Institute of Technology, Pasadena, CA 91125, USA}

\author[0000-0003-3504-5316]{Benjamin J.\ Fulton}
\affiliation{Cahill Center for Astronomy $\&$ Astrophysics, California Institute of Technology, Pasadena, CA 91125, USA}
\affiliation{IPAC-NASA Exoplanet Science Institute, Pasadena, CA 91125, USA}

\author[0000-0003-1125-2564]{Ashley Chontos}
\affiliation{Institute for Astronomy, University of Hawai$'$i, Honolulu, HI 96822, USA}
\affiliation{NSF Graduate Research Fellow}

\author[0000-0003-0800-0593]{Justin R.\ Crepp}
\affiliation{Department of Physics, University of Notre Dame, Notre Dame, IN, 46556, USA}

\author[0000-0002-4297-5506]{Paul A.\ Dalba}
\affiliation{Department of Earth and Planetary Sciences, University of California, Riverside, CA 92521, USA}
\affiliation{NSF Astronomy $\&$ Astrophysics Postdoctoral Fellow}

\author[0000-0003-4155-8513]{Gregory W.\ Henry}
\affiliation{Center of Excellence in Information Systems, Tennessee State University, Nashville, TN 37209 USA}

\author[0000-0002-7084-0529]{Stephen R.\ Kane}
\affiliation{Department of Earth and Planetary Sciences, University of California, Riverside, CA 92521, USA}

\author[0000-0003-0967-2893]{Erik A.\ Petigura}
\affiliation{Department of Physics $\&$ Astronomy, University of California Los Angeles, Los Angeles, CA 90095, USA}

\author[0000-0002-3725-3058]{Lauren M.\ Weiss}
\affiliation{Department of Physics, University of Notre Dame, Notre Dame, IN 46556, USA}

\author[0000-0001-6160-5888]{Jason T.\ Wright}
\affiliation{Department of Astronomy and Astrophysics, The Pennsylvania State University, University Park, PA 16802, USA}
\affiliation{Center for Exoplanets and Habitable Worlds, The Pennsylvania State University, University Park, PA 16802, USA}
\affiliation{Penn State Extraterrestrial Intelligence Center, The Pennsylvania State University, University Park, PA 16802, USA}

\begin{abstract}

We use a high-precision radial velocity survey of FGKM stars to study the conditional occurrence of two classes of planets: close-in small planets (0.023--1 au, 2--30 $\mearth$) and distant giant planets (0.23--10 au, 30--6000 $\mearth$). We find that $41^{+15}_{-13}\%$ of systems with a close-in, small planet also host an outer giant, compared to $17.6^{+2.4}_{-1.9}\%$ for stars irrespective of small planet presence. This implies that small planet hosts may be enhanced in outer giant occurrence compared to all stars with $1.7\sigma$ significance. Conversely, we estimate that $42^{+17}_{-13}\%$ of cold giant hosts also host an inner small planet, compared to $27.6^{+5.8}_{-4.8}\%$ of stars irrespective of cold giant presence. We also find that more massive and close-in giant planets are not associated with small inner planets. Specifically, our sample indicates that small planets are less likely to have outer giant companions more massive than approximately 120 $\mearth$ and within 0.3--3 au than to have less massive or more distant giant companions, with $\sim$2.2$\sigma$ confidence. This implies that massive gas giants within 0.3--3 au may suppress inner small planet formation. Additionally, we compare the host-star metallicity distributions for systems with only small planets and those with both small planets and cold giants. In agreement with previous studies, we find that stars in our survey that only host small planets have a metallicity distribution that is consistent with the broader solar-metallicity-median sample, while stars that host both small planets and gas giants are distinctly metal-rich with $\sim$2.3$\sigma$ confidence.
\end{abstract}

\keywords{exoplanets}

\section{Introduction} \label{sec:intro}

The relationship between small, close-in planets and outer giant companions reveals much about planet formation. Gas giant interactions with protoplanetary disks can create low-density gaps that halt the inward drift of gas and solids, possibly suppressing the formation of close-in small planets \citep{Lin86, Moriarty15, Ormel17}. It is also possible that warm or eccentric giants disrupt the growth of of small planets by pebble or planetesimal accretion, or destabilize the orbits of nascent small planet cores, as predicted by synthetic population studies \citep{Bitsch20, Schlecker20}. These phenomena would lead to a population of small planets without outer giant companions, or an absence of companions within a certain range of mass, semi-major axis, and eccentricity.

On the other hand, the same stellar properties that facilitate giant planet formation, such as high metallicity \citep{Fischer05}, may also enhance small planet formation.  If higher metallicity stars were more likely to form super-Earths, we would expect to see a metallicity dependence in their observed occurrence rate regardless of the presence or absence of an outer companion. \cite{Petigura18} analyzed the metallicity distribution of \textit{Kepler} stars and planet hosts and found that warm sub-Neptune (1.7--4 $\rearth$) occurrence is weakly correlated with host-star metallicity, doubling from -0.4 dex to +0.4 dex with $\sim$2$\sigma$ significance. However, \cite{Moe19} and \cite{Kutra20} later found that this correlation disappears if one decorrelates against the metallicity dependence of close binaries, which do not host short-period planets. This leaves open the possibility that the giant planets formed in metal-rich disks might directly facilitate small planet formation via their dynamical impact on the protoplanetary disk structure \citep[e.g.,][]{Hasegawa11, Buchhave14}.

We can explore the tension between these ideas by using exoplanet surveys to measure the conditional probability that gas giant hosts also host at least one inner small planet, and comparing that value to the overall occurrence rates of close-in small planets and distant giant planets, beyond roughly 0.3 au. If inner small planet companions to cold giants are rarities compared to the broader sample of small planets, then we can deduce that giant planets in a certain mass and semi-major axis range suppress small planet formation. Conversely, if small planet companions to cold giants are common, this implies that disks that form cold gas giants also provide favorable conditions for small planet formation, or that cold giants actively facilitate small planet formation.

Recently, \cite{Zhu18} and \cite{Bryan19} independently used samples of stars with known super-Earths, most of which have masses less than 10 $\mearth$ , to directly estimate the fraction of super-Earth hosts that have outer gas giant companions. Furthermore, each analysis used Jupiter-analog occurrence rates from \cite{Wittenmyer16} and \cite{Rowan16} to infer the fraction of cold giants that host inner super-Earths. \cite{Bryan19} found that $102^{+34}_{-51}\%$ of stars with Jupiter analogs (3--7 au, 0.3--13 \mjup) also host an inner super-Earth, while \cite{Zhu18} reported $90\pm 20\%$ for the same measurement. Both studies predicted that nearly all Jupiter analogs host inner small planets, with high uncertainties in conditional probability due to the indirect nature of this Bayesian inference. So far, no study has directly measured the rate at which gas giants are accompanied by inner small planets. To make this measurement, we need a large sample of cold Jupiters with RV data sets that are also sensitive to the presence of small inner planets, which requires long-baseline radial velocity (RV) observations.  The measurement must also be sensitive to small inner planets, which requires high-cadence radial velocities or coverage by a photometric transit survey. The former is a costly undertaking that can only be done over many nights of ground-based RV observing, and the latter can only detect planets with edge-on or nearly edge-on orbits.

The California Legacy Survey \citep[CLS;][]{Rosenthal21} is uniquely well-suited for this measurement. As a blind RV survey of 719 stars over three decades, it produced a sample that is appropriate for a variety of occurrence measurements, is rich in cold giants, and contains enough stars with high-cadence observations that we have some sensitivity in the small-planet regime. In this paper, we leverage this survey to explore the relationship between close-in small planets, which we limit to 0.03--1 au and 2--30 $\mearth$, and outer giant companions, which we constrain in two different ways defined below. In Section 2, we review the star and planet catalog of the California Legacy Survey. In Section 3, we describe our methods for computing planet occurrence. In Section 4, we present our results. In Section 5, we discuss our findings and their context.

\section{Survey Review}

\begin{figure*}[ht!]
\begin{center}
\includegraphics[width = \textwidth]{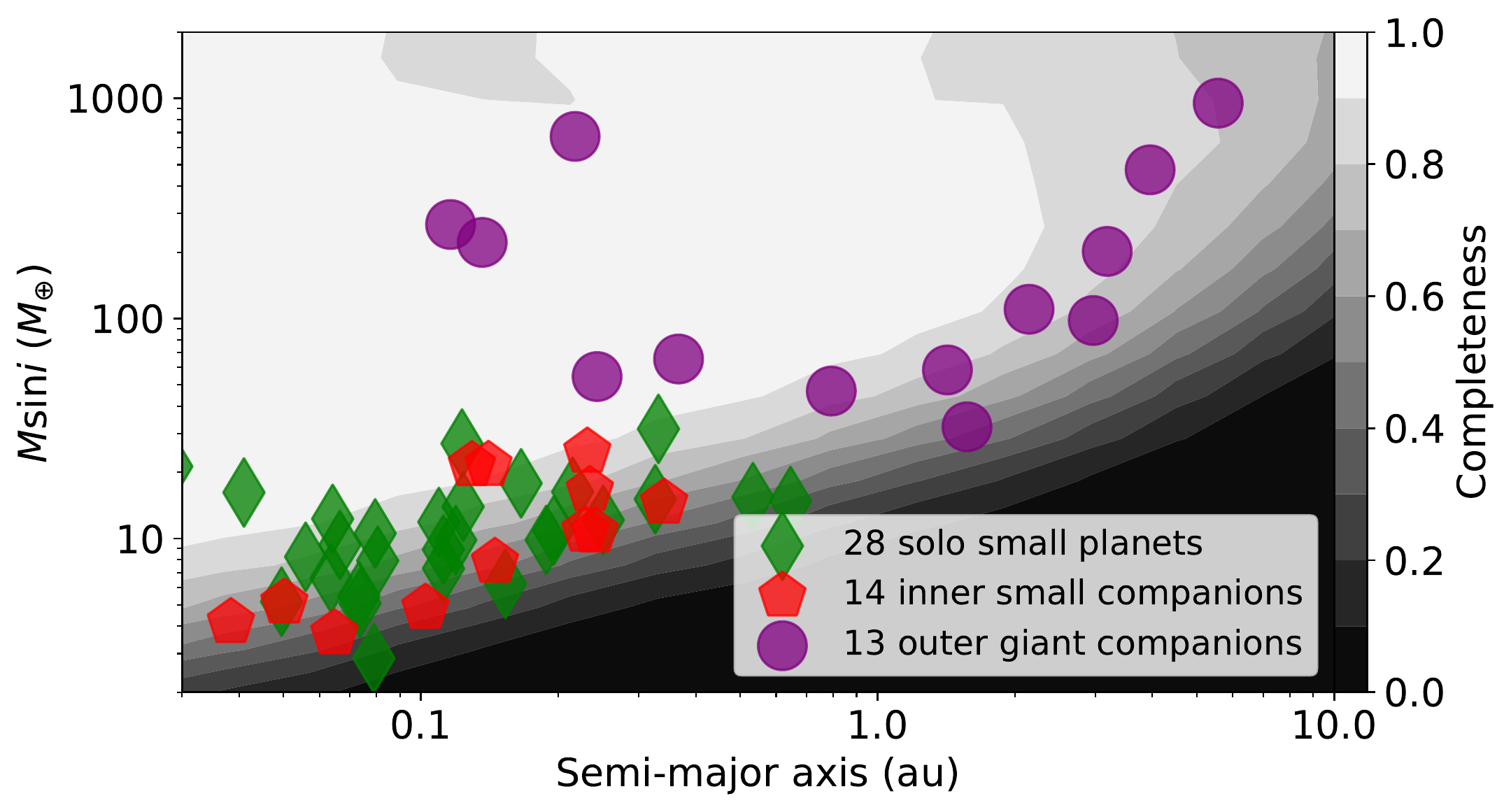}
\caption{Minimum mass (\msini) versus semi-major axis values for small ($\msini < 30 \mearth$) planets, and their giant outer companions, in the CLS catalog. Diamonds are small planets without outer giants, pentagons are small planets with outer giants, and circles are outer giants. Contours are the completeness map for small planet hosts.}
\label{fig:catalog}
\end{center}
\end{figure*}

The California Legacy Survey is a sample of 719 RV-observed FGKM stars and their associated planets created to provide a stellar and planetary catalog for occurrence studies \citep{Rosenthal21}. We approximated a quantifiably complete survey by selecting 719 stars that were observed by the California Planet Search (CPS) \citep{Howard10} and originally chosen without bias towards a higher or lower than average likelihood of hosting planets. We took our first observations in 1988 with the Lick-Hamilton spectrograph \citep{Fischer14}, and our latest observations in 2020 with the Keck-HIRES and the Lick-APF spectrographs. Our typical observational baseline is 22 yr, and our typical RV precision is 2 m s$^{-1}$. We used an automated and repeatable iterative periodogram method to search for planet candidates, implemented in the open-source package \texttt{RVSearch} \citep{Rosenthal21}, and performed uniform vetting to identify false positives. This left us with 178 planets in our sample, 43 planets with $\msini\ < 30 \mearth$, and 135 planets with $\msini\ \geq\ 30 \mearth$. Figure \ref{fig:catalog} shows our sample of small close-in planets and outer giant companions.

Our stellar sample has a median metallicity of 0.0 [Fe/H], a median stellar mass equal to 1.0 $M_{\odot}$, and a small number of evolved stars. These are reasonable heuristics for verifying that we successfully constructed a planet-blind occurrence survey, since a bias towards known giant planet hosts could manifest as a metal-rich sample \citep{Fischer05}, a particularly massive sample, or an excess of evolved stars \citep{Johnson11}.

Since the CLS drew from the CPS RV catalog, our sample encompasses stars from several Keck-HIRES occurrence surveys, including \cite{Cumming08} and \cite{Howard10_Science}. We refer the reader to \cite{Rosenthal21} for the full star and planet catalog, as well as details of the planet search and completeness characterization.

\begin{longtable*}{llrrrrr}
\caption{Small Planet Sample} \\
\toprule 
\midrule 

Name & \msini\ [$\mearth$] & $a$ [AU] & Discovery Ref. \\ 
\toprule 
\endfirsthead 
\caption[]{Small Planet Sample (Continued)} \\
\toprule 
\midrule 
Name & \msini\ [$\mearth$] & $a$ [AU] & Discovery Ref. \\ 
\toprule 
\endhead 
HD 107148 b & $19.9^{+3.1}_{-3.1}$ & $0.1407^{+0.0018}_{-0.0019}$ & \cite{Butler06} & \\ 
HD 115617 b & $16.1^{+1.1}_{-1.2}$ & $0.2151^{+0.0028}_{-0.0029}$ & \cite{Vogt10} & \\ 
HD 115617 c & $5.11^{+0.53}_{-0.51}$ & $0.04956^{+0.00065}_{-0.00067}$ & \cite{Vogt10} & \\ 
HD 11964A b & $24.4^{+2.0}_{-2.0}$ & $0.2315^{+0.0021}_{-0.0022}$ & \cite{Butler06} & \\ 
HD 1326 b & $5.43^{+0.42}_{-0.42}$ & $0.0732^{+0.00047}_{-0.00048}$ & \cite{Howard14} & \\ 
HD 141004 b & $13.6^{+1.5}_{-1.4}$ & $0.1238^{+0.002}_{-0.002}$ & \cite{Rosenthal21} & \\ 
HD 1461 b & $6.6^{+0.61}_{-0.56}$ & $0.0636^{+0.00095}_{-0.00099}$ & \cite{Rivera10} & \\ 
HD 1461 c & $7.07^{+0.88}_{-0.90}$ & $0.1121^{+0.0017}_{-0.0017}$ & \cite{Diaz16} & \\ 
HD 147379A b & $30.7^{+3.7}_{-3.8}$ & $0.3315^{+0.0024}_{-0.0024}$ & \cite{Reiners18} & \\ 
HD 156668 b & $5.03^{+0.42}_{-0.42}$ & $0.05024^{+0.00051}_{-0.00052}$ & \cite{Howard11} & \\ 
HD 164922 b & $14.3^{+1.1}_{-1.1}$ & $0.3411^{+0.0039}_{-0.0039}$ & \cite{Butler06} & \\ 
HD 164922 c & $10.53^{+0.98}_{-0.98}$ & $0.2292^{+0.0026}_{-0.0027}$ & \cite{Fulton16} & \\ 
HD 164922 d & $4.73^{+0.66}_{-0.66}$ & $0.1023^{+0.0012}_{-0.0012}$ & \cite{Rosenthal21} & \\ 
HD 168009 b & $9.5^{+1.2}_{-1.2}$ & $0.1192^{+0.0017}_{-0.0018}$ & \cite{Rosenthal21} & \\ 
HD 190360 b & $21.44^{+0.85}_{-0.84}$ & $0.1294^{+0.0017}_{-0.0017}$ & \cite{Naef03} & \\ 
HD 192310 b & $14.3^{+2.0}_{-1.9}$ & $0.3262^{+0.0036}_{-0.0037}$ & \cite{Howard11} & \\ 
HD 216520 b & $10.4^{+1.1}_{-1.2}$ & $0.1954^{+0.0025}_{-0.0025}$ & \cite{Burt21} & \\ 
HD 219134 b & $16.41^{+1.00}_{-0.95}$ & $0.2345^{+0.0027}_{-0.0027}$ & \cite{Vogt15} & \\ 
HD 219134 c & $4.12^{+0.33}_{-0.34}$ & $0.03838^{+0.00044}_{-0.00044}$ & \cite{Vogt15} & \\ 
HD 219134 d & $7.73^{+0.73}_{-0.69}$ & $0.1453^{+0.0017}_{-0.0016}$ & \cite{Vogt15} & \\ 
HD 219134 e & $3.57^{+0.43}_{-0.45}$ & $0.06466^{+0.00074}_{-0.00073}$ & \cite{Vogt15} & \\ 
HD 285968 b & $9.1^{+1.4}_{-1.4}$ & $0.06649^{+0.00043}_{-0.00043}$ & \cite{Forveille09} & \\ 
HD 42618 b & $15.2^{+1.8}_{-1.8}$ & $0.5337^{+0.0088}_{-0.0091}$ & \cite{Fulton16} & \\ 
HD 45184 b & $11.9^{+1.3}_{-1.2}$ & $0.0641^{+0.0011}_{-0.0011}$ & \cite{Udry19} & \\ 
HD 45184 c & $10.9^{+1.8}_{-1.8}$ & $0.1095^{+0.0018}_{-0.0018}$ & \cite{Udry19} & \\ 
HD 69830 b & $10.26^{+0.69}_{-0.64}$ & $0.0794^{+0.0012}_{-0.0012}$ & \cite{Lovis06} & \\ 
HD 69830 c & $9.86^{+0.97}_{-0.94}$ & $0.1882^{+0.0029}_{-0.0029}$ & \cite{Lovis06} & \\ 
HD 69830 d & $14.1^{+1.7}_{-1.8}$ & $0.645^{+0.01}_{-0.01}$ & \cite{Lovis06} & \\ 
HD 75732 b & $9.37^{+0.43}_{-0.43}$ & $0.01583^{+0.00024}_{-0.00024}$ & \cite{Butler97} & \\ 
HD 7924 b & $8.23^{+0.45}_{-0.44}$ & $0.05595^{+0.00075}_{-0.00078}$ & \cite{Fulton15} & \\ 
HD 7924 c & $8.83^{+0.63}_{-0.59}$ & $0.1121^{+0.0015}_{-0.0016}$ & \cite{Fulton15} & \\ 
HD 7924 d & $6.1^{+0.68}_{-0.65}$ & $0.1532^{+0.0021}_{-0.0021}$ & \cite{Fulton15} & \\ 
HD 90156 b & $11.8^{+2.0}_{-1.9}$ & $0.2509^{+0.0036}_{-0.0037}$ & \cite{Mordasini11} & \\ 
HD 95735 b & $18.0^{+2.9}_{-2.6}$ & $3.1^{+0.13}_{-0.11}$ & \cite{Diaz19} & \\ 
HD 97101 b & $10.2^{+1.3}_{-1.2}$ & $0.2403^{+0.0017}_{-0.0017}$ & \cite{Dedrick21} & \\ 
HD 97658 b & $7.85^{+0.57}_{-0.55}$ & $0.0805^{+0.0010}_{-0.0011}$ & \cite{Howard11} & \\ 
HD 99492 b & $26.7^{+1.9}_{-1.9}$ & $0.1231^{+0.0014}_{-0.0015}$ & \cite{Marcy05} & \\ 
GL 687 b & $17.6^{+1.5}_{-1.5}$ & $0.1658^{+0.0012}_{-0.0012}$ & \cite{Burt14} & \\ 
HIP 74995 b & $16.22^{+0.62}_{-0.61}$ & $0.04099^{+0.00042}_{-0.00044}$ & \cite{Bonfils05} & \\ 
HIP 74995 c & $5.06^{+0.69}_{-0.69}$ & $0.07359^{+0.00076}_{-0.00078}$ & \cite{Mayor09} & \\ 
HIP 57087 b & $21.22^{+0.70}_{-0.69}$ & $0.02849^{+0.0002}_{-0.0002}$ & \cite{Butler04} & \\ 
GL 876 b & $5.86^{+0.50}_{-0.49}$ & $0.02183^{+0.00018}_{-0.00019}$ & \cite{Marcy01} & \\ 
\bottomrule 
\end{longtable*}

\section{Methods}

\subsection{Occurrence model}

The primary goal of this work is to measure planet occurrence, particularly of small close-in planets and cold gas giants. Many studies of RV or transit surveys use the intuitive occurrence measurement method known as ``inverse detection efficiency" \citep{Howard12, Petigura13b}. According to this procedure, one estimates occurrence in a region of parameter space by counting up the planets found in that region, with each planet weighted by the search completeness in that region. We measured the search completeness map of our survey by injecting many synthetic signals into each dataset, and computing the fraction of signals in a given region that are recovered by our search algorithm, \texttt{RVSearch}. Inverse detection efficiency as defined in \cite{Foreman-Mackey14} is actually a specific case of a Poisson likelihood method, in which one models an observed planet catalog as the product of an underlying Poisson process and empirical completeness map.

Following the analysis in \cite{Fulton21}, we used the Poisson likelihood method to model the occurrence of planets. Given a population of observed planets with orbital and \msini\ posteriors $\{\bm{\omega}\}$, and associated survey completeness map $Q(\bm{\omega})$, and assuming that our observed planet catalog is generated by a set of independent Poisson process draws, we can evaluate a Poisson likelihood for a given occurrence model $\Gamma(\bm{\omega} | \bm{\theta})$, where $\bm{{\theta}}$ is a vector parameterizing the rates of the Poisson process. The observed occurrence $\hat{\Gamma}(\bm{\omega} | \bm{\theta})$ of planets in our survey can be modeled as the product of the measured survey completeness and some underlying occurrence model,

\begin{equation}
\hat{\Gamma}(\bm{\omega} | \bm{\theta}) = Q(\bm{\omega})\Gamma(\bm{\omega} | \bm{\theta}). \\
\end{equation}

The Poisson likelihood for an observed population of objects is

\begin{equation}
\mathcal{L} = e^{-\int \hat{\Gamma}(\bm{\omega} | \bm{\theta}) \,d\bm{\omega}} \prod_{k=1}^{K} \hat{\Gamma}(\bm{\omega}_k | \bm{\theta}),\\
\end{equation}
where $K$ is the number of observed objects, and $\bm{\omega}_k$ is the $k$th planet's orbital parameter vector. The Poisson likelihood can be understood as the product of the probability of detecting an observed set of objects (the product term in Equation 2) and the probability of observing no additional objects in the considered parameter space (the integral over parameter space). Equations 1 and 2 serve as the foundation for our occurrence model, but do not take into account uncertainty in our measurements of planetary orbits and minimum masses. In order to do this, we use Markov Chain Monte Carlo methods to empirically sample the orbital posteriors of each system \citep{DFM13, Fulton18, Rosenthal21}. We can hierarchically model the orbital posteriors of each planet in our catalog by summing our occurrence model over many posterior samples for each planet. The hierarchical Poisson likelihood is therefore approximated as

\begin{equation}
\mathcal{L}\approx e^{-\int \hat{\Gamma}(\bm{\omega} | \bm{\theta}) \,d\bm{\omega}} \prod_{k=1}^{K} \frac{1}{N_k} \sum_{n=1}^{N_k} \frac{\hat{\Gamma}(\bm{\omega}_k^n | \bm{\theta})}{p(\bm{\omega}_k^n | \bm{\alpha})},\\
\end{equation}
where $N_k$ is the number of posterior samples for the $k$th planet in our survey, and $\bm{\omega}_k^n$ is the $n$th sample of the $k$th planet's posterior. $p(\bm{\omega} | \bm{\alpha})$ is our prior on the individual planet posteriors. We placed uniform priors on ln($M$sin$i$) and ln($a$). We used \texttt{emcee} to sample our hierarchical Poisson likelihood, and placed uniform priors on $\theta$.

\subsection{Approach to planet multiplicity}

We want to evaluate the link between the presence of any inner small planets and the presence of any cold gas giants. Therefore, for all combinations of the presence or absence of these two planet types, we are interested in estimating the probability that a star hosts at least one planet. This quantity is distinct from the number of planets per star, both because many stars host more than one small planet \citep{Howard12, Fang12, He20} and because the probability of hosting at least one planet must be less than 1. We attempt to resolve this issue with two constraints on our model. First, we place a hard-bound prior on the integrated occurrence rate, so that it has an upper limit of one planet per star. Second, in the case of planetary systems that contain multiple detected planets in the class of interest, we only count the planet that was first detected by our search algorithm. We also report the expected number of planets per star in Table \ref{tab:conditionals}, by including all companions in multi-planet systems.

\begin{figure}[ht!]
\includegraphics[width = 0.5\textwidth]{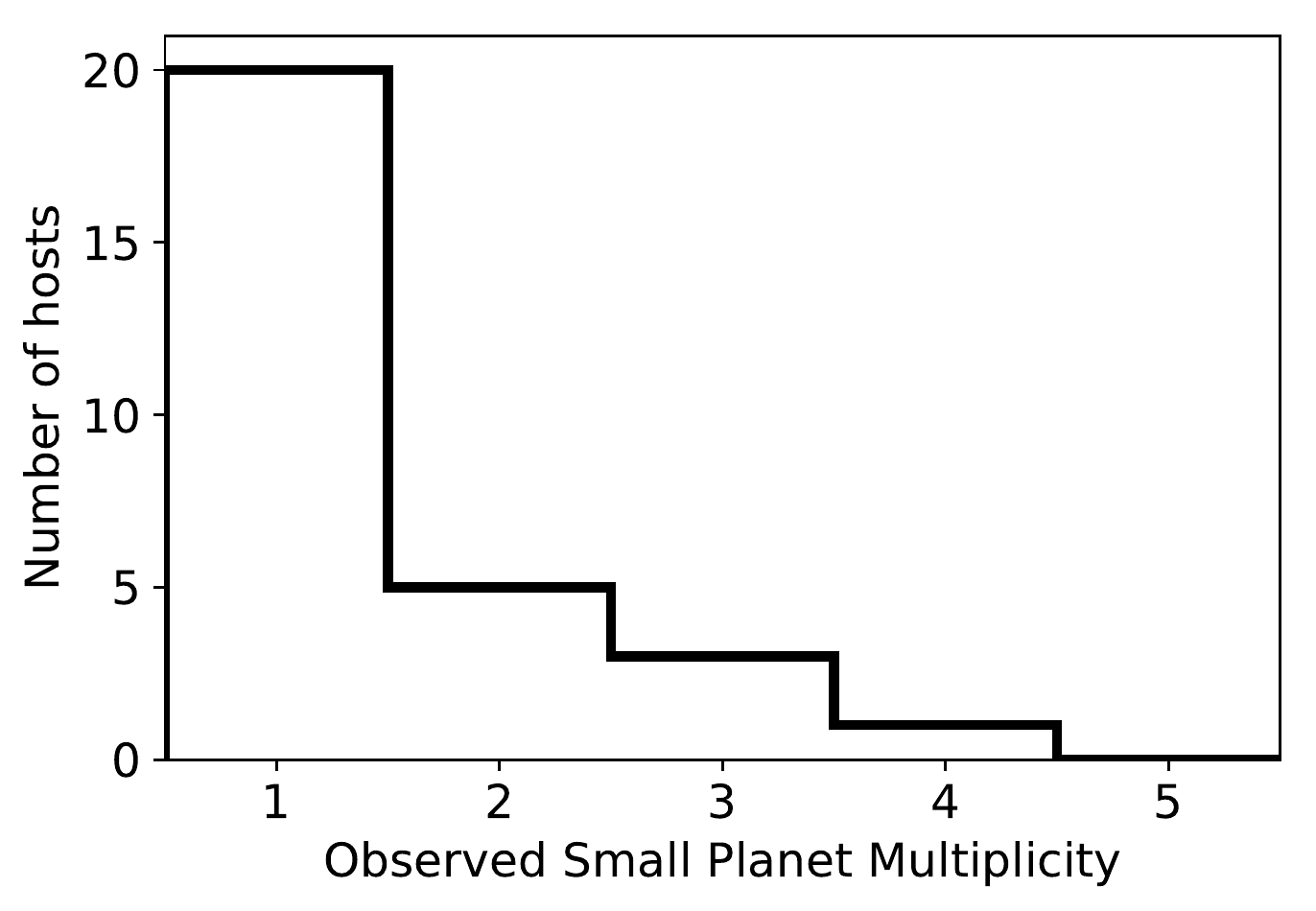}
\caption{A histogram of observed small planet multiplicity in our sample. This is not corrected for search completeness, so it should only be interpreted as the multiplicity of detected planets, not as the underlying multiplicity distribution. There are 719 total stars in the CLS, around 29 of which we have detected small planets.}
\label{fig:multiplicity}
\end{figure}

The resulting estimate of the probability that a star hosts at least one planet depends on the search completeness in the mass and semi-major axis range of each individual planet. This biases our sample towards planets with greater RV semi-amplitudes, which tend to be closer-in and higher-mass. These planets are in higher-completeness regions and therefore will usually be detected first by iterative search algorithms. Figure \ref{fig:multiplicity} shows the observed multiplicity of the detected small planets in our sample. Note that this distribution is not corrected for search completeness, so it cannot be interpreted as the true underlying multiplicity distribution. Rather, it is showing how many multi-planet systems we detect with respect to systems where we only detect one small planet.

\section{Results} \label{sec:results}

\subsection{Absolute and conditional occurrence rates}

\begin{figure*}[ht!]
\begin{center}
\includegraphics[width = 0.495\textwidth]{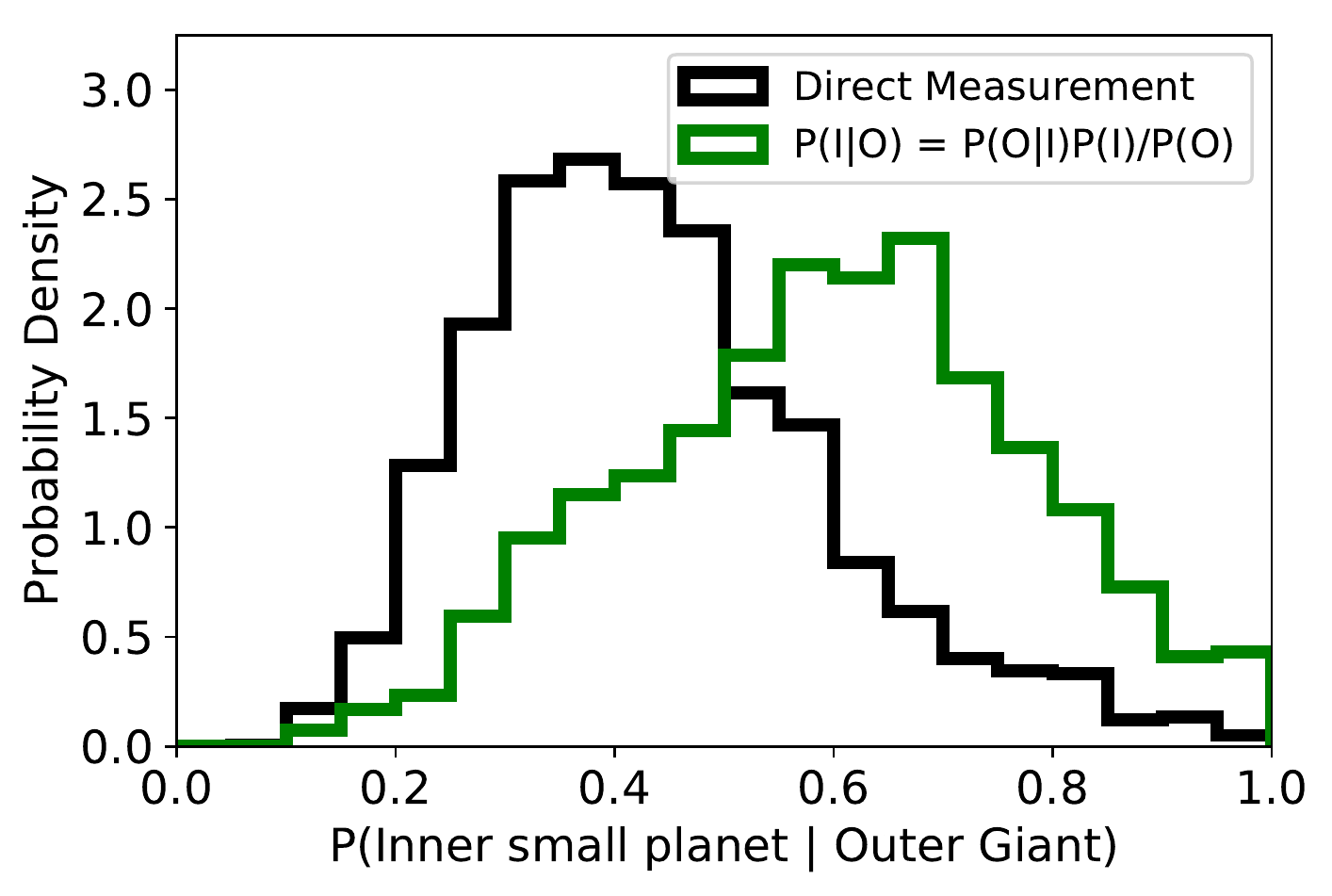}
\includegraphics[width = 0.495\textwidth]{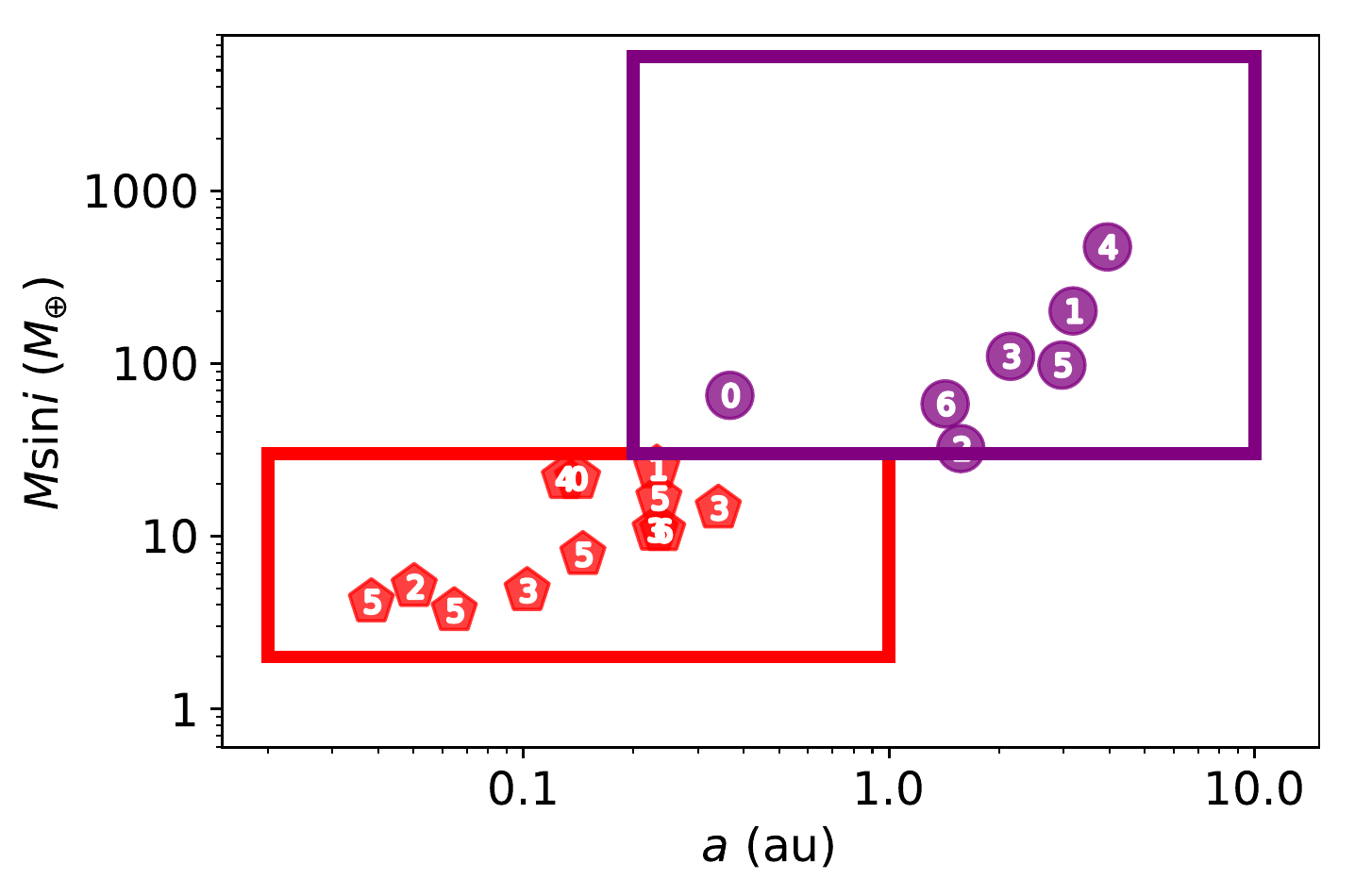}
\caption{Left: Two measurements of the conditional occurrence of inner small planets given the presence of an outer gas giant. The black distribution is our direct measurement, while the green distribution uses Bayes Theorem to infer it from other measurements. Right: Our sample of planet pairs with small planets within the region of interest, with our inner small planet box outlined in red and our outer giant box outlined in purple. We assign a number to each planetary system and label individual planets accordingly.}
\label{fig:bayes_simple}
\end{center}
\end{figure*}

\begin{figure*}[ht!]
\begin{center}
\includegraphics[width = 0.495\textwidth]{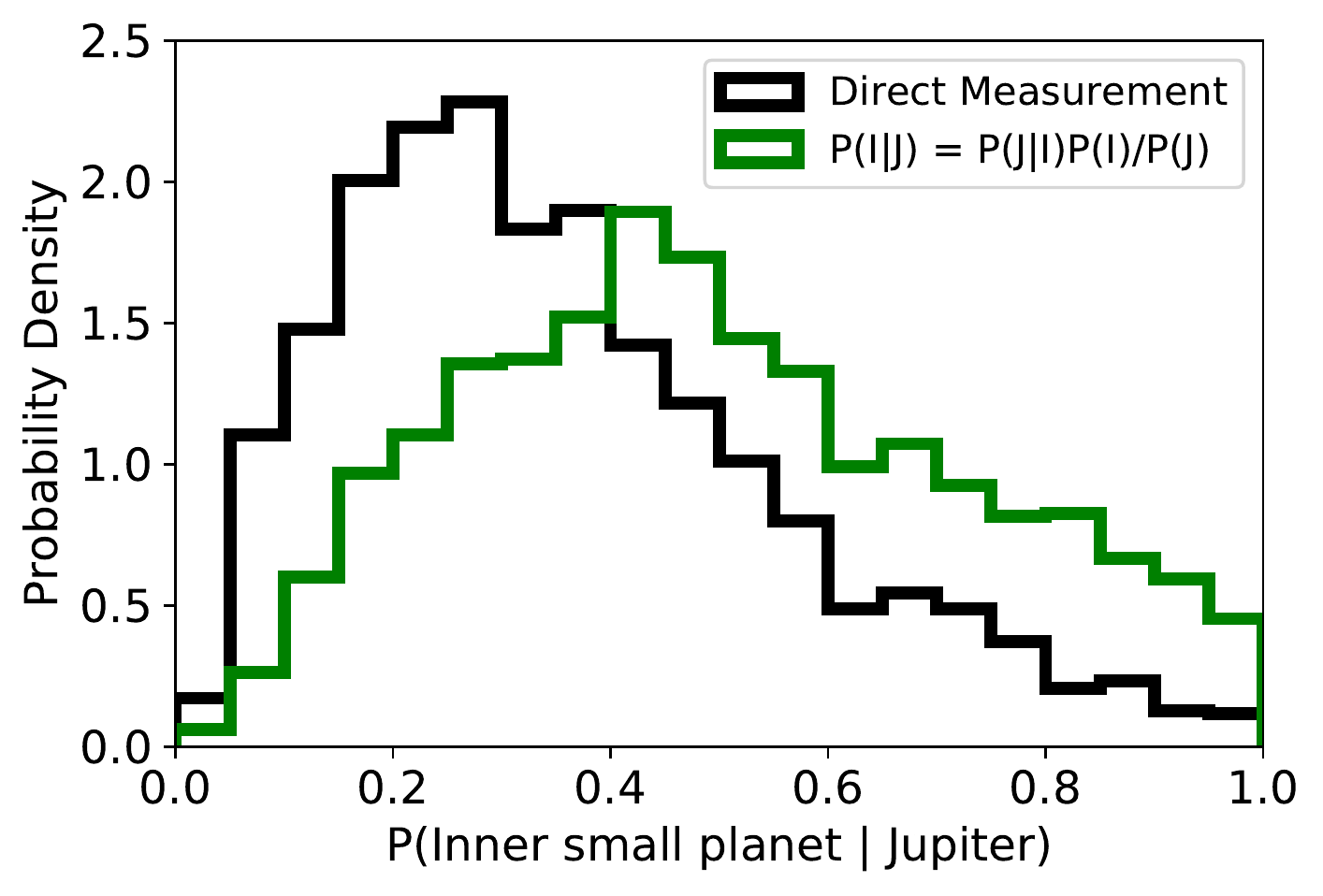}
\includegraphics[width = 0.495\textwidth]{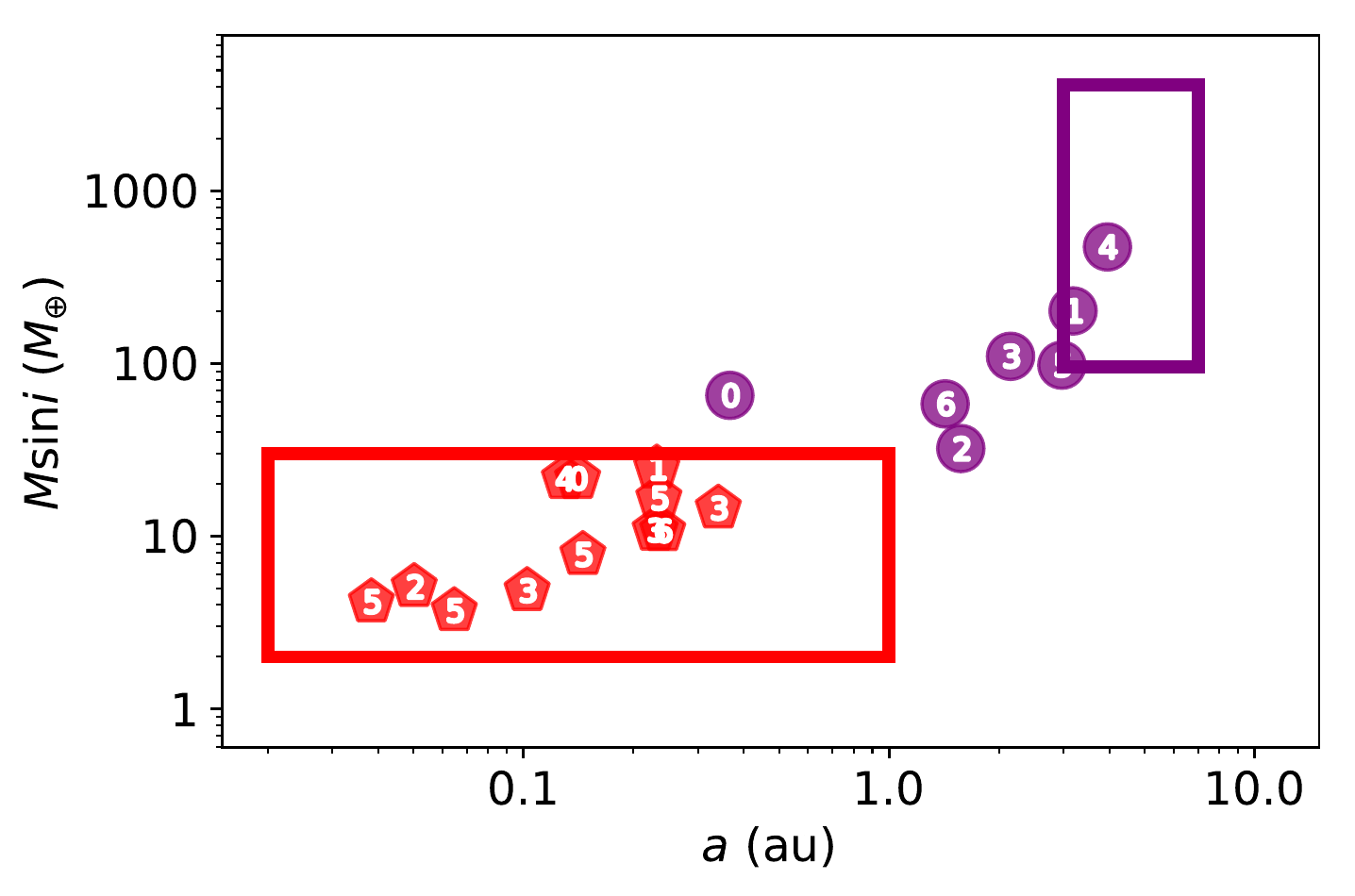}
\caption{Same as Figure \ref{fig:bayes_simple}, but for Jupiter analogs within 3--7 au and 0.3--13 \mjup\  instead of the broader giant population.}
\label{fig:bayes_jovian}
\end{center}
\end{figure*}

Using our occurrence methodology, we measured a set of distinct occurrence probabilities for the CLS sample. Specifically, we computed the absolute probability of hosting a small close-in planet, $P(I)$; the absolute probability of hosting a cold gas giant, $P(O)$; The probability of hosting a cold gas giant given the presence of a small close-in planet, $P(O|I)$; and the probability of hosting a small close-in planet given the presence of a cold gas giant, $P(I|O)$. In each case, we used our approach to multiplicity to link $P(\bm{\omega} | \bm{\theta})$ with $\Gamma(\bm{\omega} | \bm{\theta})$. We define the $I$ range as 0.02--1 au and 2--30 $\mearth$. We define the $O$ range in two ways: broadly, with 30--6000 $\mearth$ and 0.23--10 au; and to only encompass Jupiter analogs as defined in \cite{Wittenmyer16} and \cite{Bryan19}, with 3--7 au and 95--4130 $\mearth$. Figure \ref{fig:bayes_simple} shows $P(I|O)$ for the broad definition of giant planets, while Figure \ref{fig:bayes_jovian} shows $P(I|O)$ for Jupiter analogs. Table \ref{tab:conditionals} reports all absolute and conditional probabilities for these populations, as well as average number of planets per star, both for broad gas giants and Jupiter analogs. It shows that $P(I)$ and $P(I|O)$ are not significantly separated from each other, at least partially because the uncertainty in our measurement of $P(I|O)$ is high. In the following subsections, we compute the significance of the separation between two probability distributions as

\begin{equation}
S = \frac{|\bar{P_2} - \bar{P_1}|}{\sqrt{\sigma_{P_2}^2 + \sigma_{P_1}^2}},\\
\end{equation}
where $\bar{P}$ is the mean of a distribution and $\sigma_{P}^2$ is its variance.

\begin{deluxetable}{lrr}
\caption{Absolute and conditional probabilities and occurrences for inner small planets, outer giants, and Jupiters.\label{tab:conditionals}
}
\tabletypesize{\large}
\tablehead{\colhead{Condition} &
\colhead{$P$(1+)} &
\colhead{$<N_P>$}}
\startdata
Inner & $0.276^{+0.058}_{-0.048}$ & $0.279^{+0.055}_{-0.053}$ \\
Outer & $0.176^{+0.024}_{-0.019}$ & $0.247^{+0.022}_{-0.023}$ \\
Jupiter & $0.072^{+0.014}_{-0.013}$ & $0.078^{+0.013}_{-0.014}$ \\
Outer$|$Inner & $0.41^{+0.15}_{-0.13}$ & $0.47^{+0.15}_{-0.12}$\\
Jupiter$|$Inner & $0.133^{+0.097}_{-0.063}$ & $0.20^{+0.12}_{-0.08}$\\
Inner$|$Outer & $0.42^{+0.17}_{-0.13}$ & $0.69^{+0.19}_{-0.19}$\\
Inner$|$Jupiter & $0.32^{+0.24}_{-0.16}$ & $0.34^{+0.24}_{-0.17}$ \\
\enddata
\end{deluxetable}

\subsection{The impact of outer giants on inner small planet occurrence, and vice versa}

Table \ref{tab:conditionals} shows that $P(I) = 0.276^{+0.058}_{-0.048}$, whereas $P(I|O) = 0.42^{+0.17}_{-0.13}$. This implies that outer giant planets, according to our broad definition, enhance the occurrence of inner small planets with $\sim$1$\sigma$ significance. Also, $P(O) = 0.176^{+0.024}_{-0.019}$, whereas $P(O|I) = 0.41^{+0.15}_{-0.13}$. This implies that inner small planets enhance the occurrence of outer giant planets with $1.65\sigma$ significance. This significance decreases when we narrow our outer companions to Jupiter analogs instead of a broad range of cold giants. In that case, $P(J|I)$ is only $0.85\sigma$ enhanced over $P(J)$, and $P(I|J)$ is not separated from $P(I)$. Additionally, whether we select a broad range of cold gas giants or a specific set of Jupiter analogs, our results rule out a 100$\%$ occurrence of small inner planets within 2--30 $\mearth$ to outer gas giants. 

\begin{figure}[ht!]
\includegraphics[width = 0.495\textwidth]{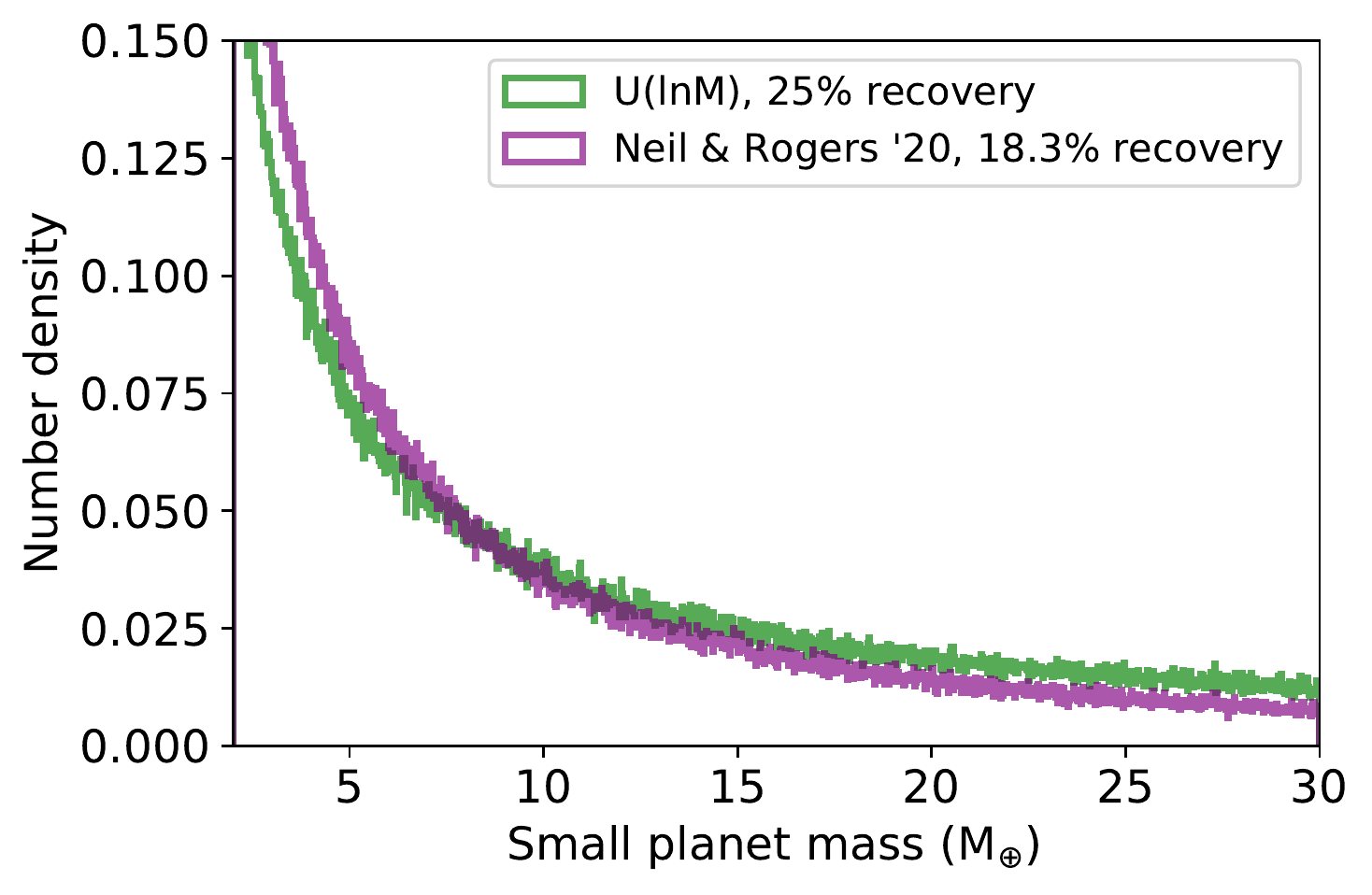}
\caption{Analytical mass distributions for small planets used in this work (green) and from \cite{Neil2020} (purple). This work assumes a uniform distribution in ln($M$). Assuming a uniform distribution in ln($a$), this leads to a 25$\%$ recovery rate in our survey of small planets within 2--30 $\mearth$ and 0.023--1 au, given our search completeness. \cite{Neil2020} fit a log-normal mixture model to a sample of \textit{Kepler} planets and found a distinct small planet component, shown here. This model leads to an 18.3$\%$ recovery rate in our survey.}
\label{fig:small_distribution}
\end{figure}

For the purposes of this work, we have assumed a small planet mass distribution that is uniform in ln($M$). Assuming a uniform distribution in ln($a$), this leads to a 25$\%$ recovery rate in our survey of small planets within 2--30 $\mearth$ and 0.023--1 au, given our search completeness. \cite{Neil2020} fit a joint mass-radius-period distribution to a sample of \textit{Kepler} planets, and found a small planet mass distribution that is approximately log-normal, with mean $\mu_{\mathrm{ln}(M/\mearth)}$ = 0.62 and $\sigma_{\mathrm{ln}(M/\mearth)}$ = 2.39. Figure \ref{fig:small_distribution} plots and compares these two distributions. We find that assuming this log-normal distribution changes our average planet recovery rate within our small planet box from $25\%$ to $18.3\%$, which would increase our corresponding occurrence rate for inner super-Earths in systems with outer gas giants from 42$\%$ to roughly 58$\%$. We conclude that our choice of mass distribution constitutes an additional source of uncertainty that is comparable to our measurement errors.

\subsection{0.3--3 au giant suppression of inner small planets}
\label{sec:suppression}

\begin{figure*}[ht!]
\begin{center}
\includegraphics[width = 0.495\textwidth]{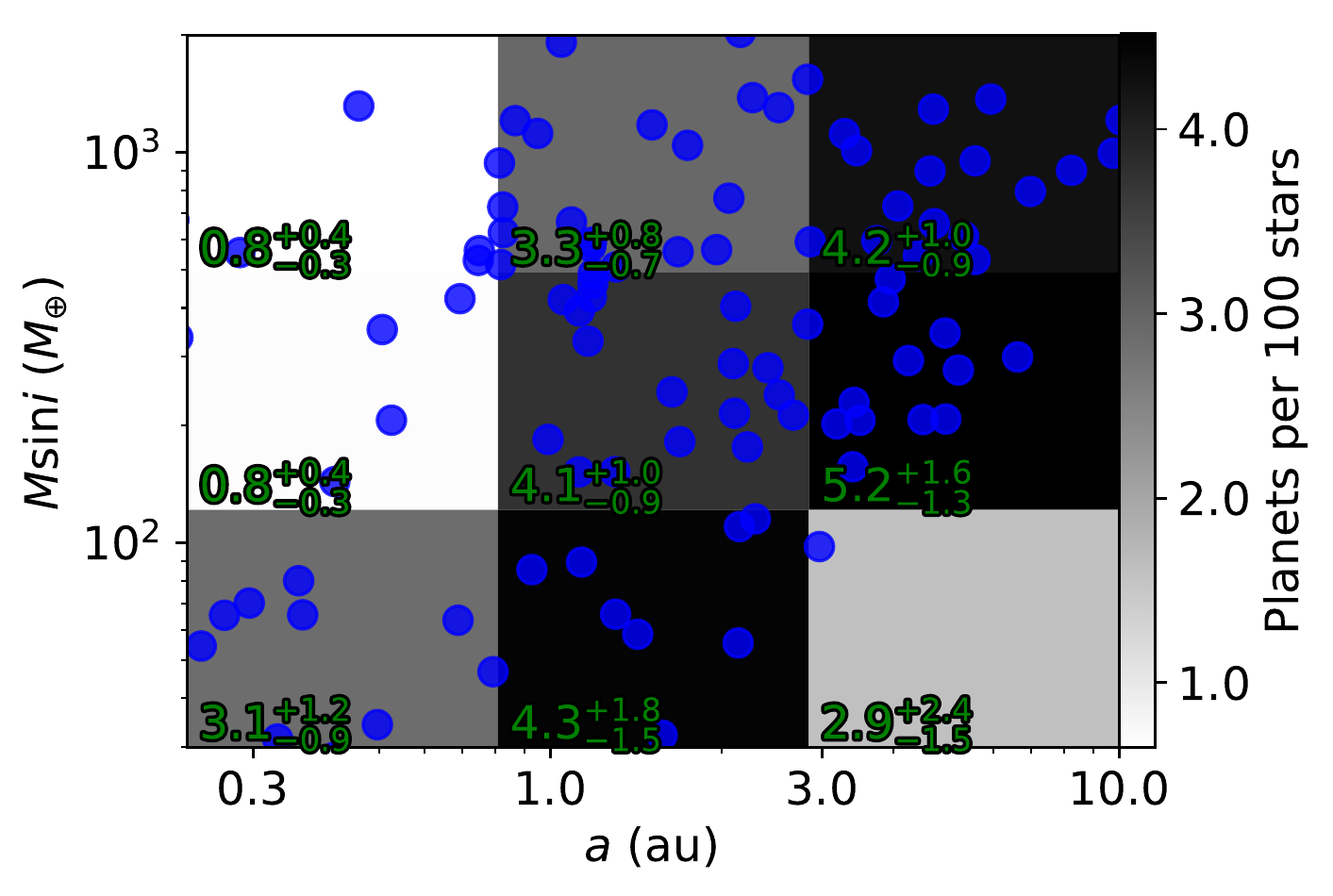}
\includegraphics[width = 0.495\textwidth]{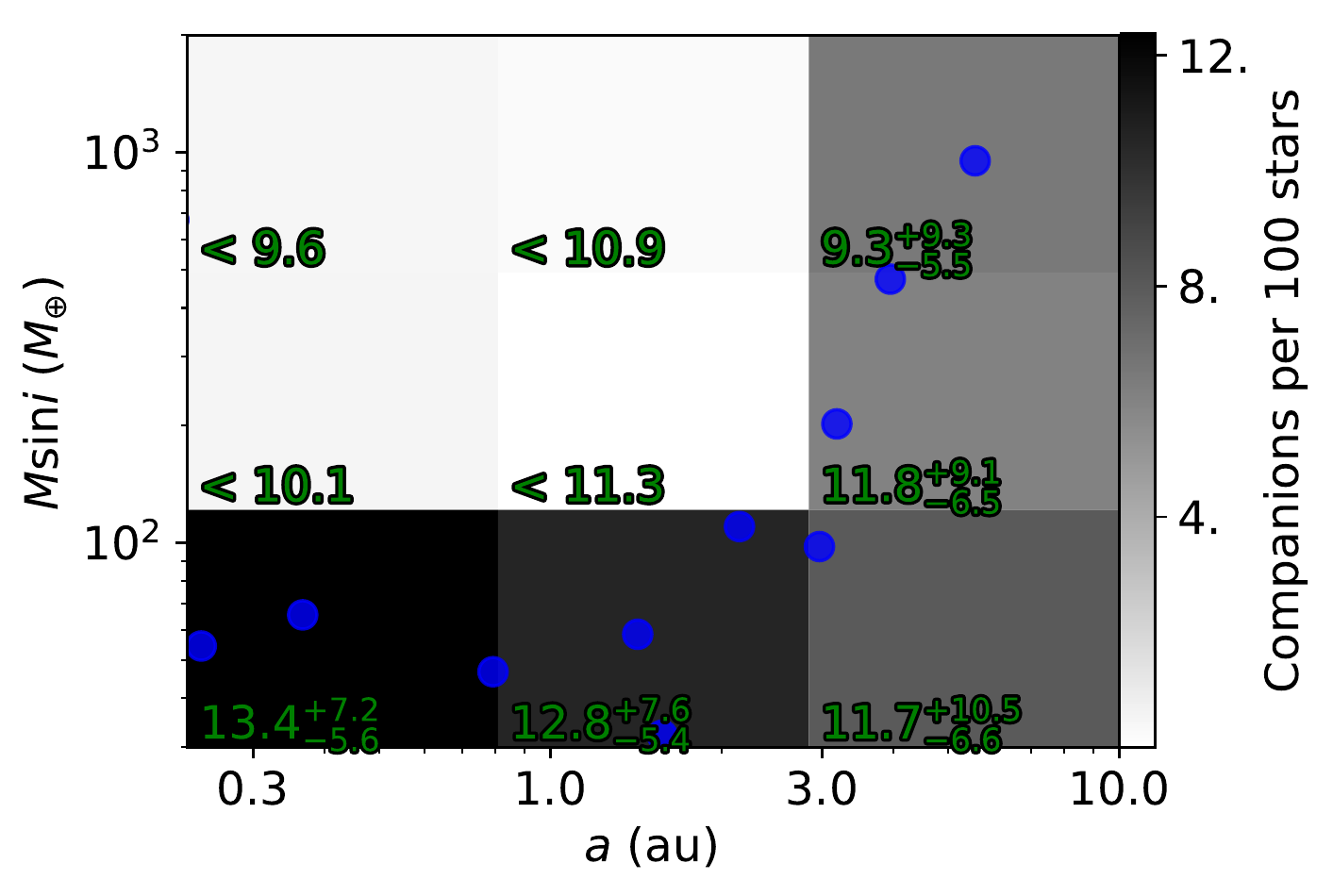}
\caption{Left: Occurrence grid for the full CLS sample of 719 stars. Cell shade and number annotation reflect the median expected number of planets per 100 stars in each bin. Empty bins show an expected upper limit on occurrence as the 84.1th percentile on the occurrence rate posterior. Right: Same, but only for the 28 hosts of detected small planets with $\msini < 30 \mearth$ and $a < 1$ AU. Note that this sample includes 55 Cnc's four cold giants, whereas our fractional analysis in Figures \ref{fig:bayes_simple} and \ref{fig:bayes_jovian} excludes 55 Cnc, since its inner ultra-short period planet is undetectable by our automated search due to our period limits. The right-hand panel shows that there is an absence of warm gas giants in our sample of detected small planet hosts.}
\label{fig:checkerboxes}
\end{center}
\end{figure*}

Figure \ref{fig:checkerboxes}, which shows occurrence grids of cold gas giants for our entire sample and for our sample of small planet hosts, provides tentative evidence that $'$lukewarm$'$ giants within roughly 0.3--3 au may suppress small planet formation. The highest-completeness region of the small planet host parameter space, within 3 au and above $\sim$120 $\mearth$, is empty, whereas there are many detected giant planets in that region without detected small companions.

We can test the significance of our absence of lukewarm giants by referring to the broader distribution of gas giants, shown in the left panel of Figure \ref{fig:checkerboxes}, and calculating the probability of drawing 10 planets (our observed outer companions) from this distribution and finding 0 within the lukewarm Jupiter region. Normalizing the occurrence map shown in Figure \ref{fig:checkerboxes}, we find a $31.1\%$ probability that a giant planet between 0.23--10 au will be found with \msini\ $>$ 120 \mjup\ and $a$ $<$ 3 au, and $68.9\%$ otherwise. We can simplify this test by using the binomial distribution to test the probability of drawing 0 out of 10 planets from a $31.1\%$ lukewarm Jupiter probability, which simplifies to $0.689^{10}=0.0241$. Thus, given our measured occurrence map for giant planets between 0.23--10 au, there is a $2.41\%$ probability of drawing 10 planets from this population and seeing 0 lukewarm Jupiters more massive than 120 \mjup\ and within 3 au. We drew this lukewarm boundary and performed this test after observing a paucity of lukewarm Jupiter companions, so it is possible that our result is biased by our sample. However, this definition of lukewarm Jupiter is physically motivated by mass and orbital separation, so it is more meaningful than a boundary arbitrarily chosen to exclude all planets.

\begin{figure}[ht!]
\includegraphics[width = 0.5\textwidth]{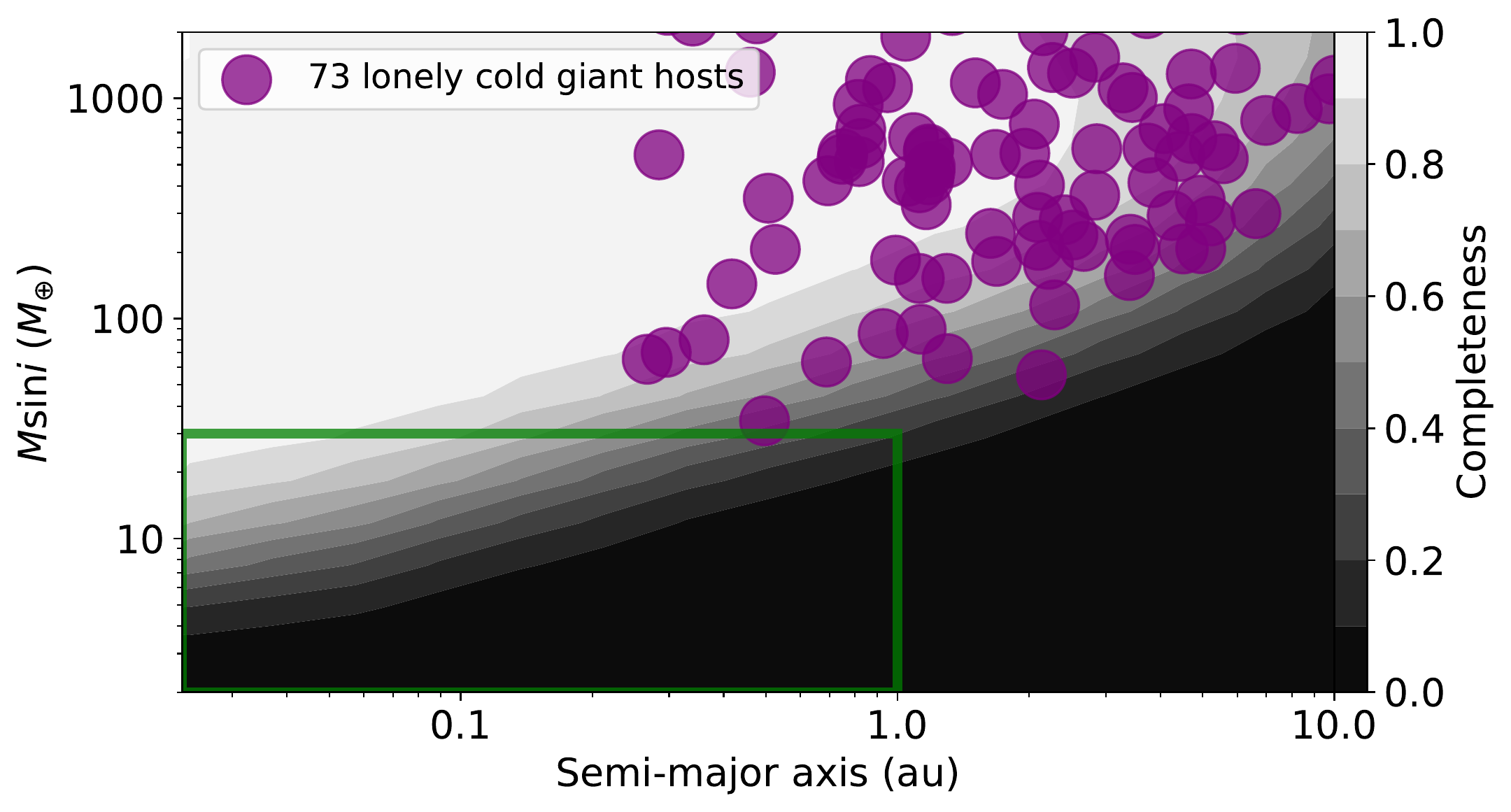}
\includegraphics[width = 0.5\textwidth]{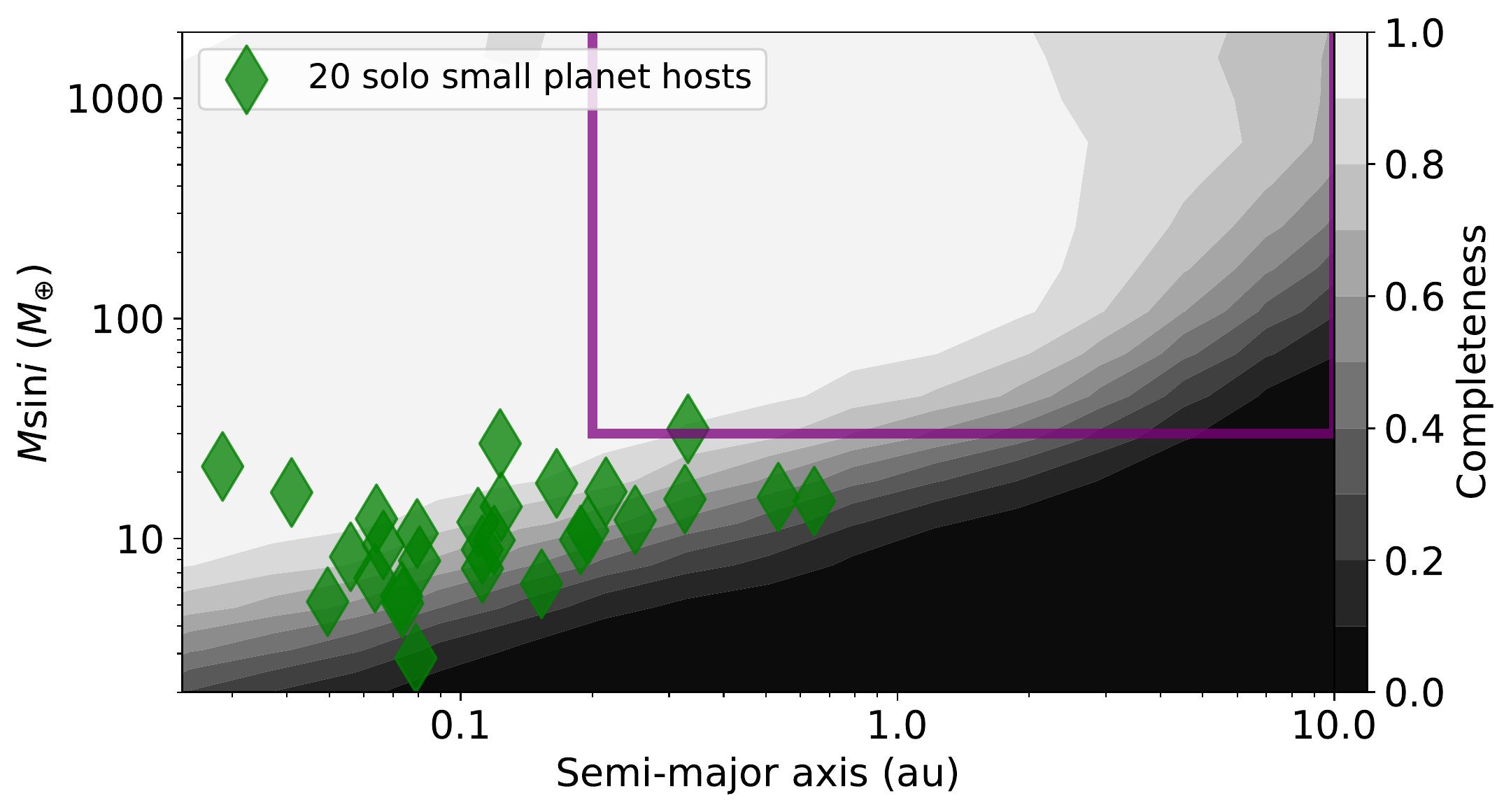}
\caption{Top: cold giant planets without detected inner small companions in the CLS sample, with associated completeness contours. The contours show that the datasets associated with these systems are somewhat sensitive to planets within 0.5 au and 30 $\mearth$ ; we have outlined our small-planet parameter space in green for context. This means that we can say with some confidence that not all of these systems harbor undetected small planets. Bottom: Small planets without detected outer companions in the CLS sample, with associated completeness contours. We have outlined our outer giant parameter space in purple for context.}
\label{fig:clean_contours}
\end{figure}

The top panel of Figure \ref{fig:clean_contours} provides additional evidence that not all giant planets beyond 0.3 au host inner small planets, since the set of systems that host cold gas giants without detected small inner planets have non-zero sensitivity to said small planets, particularly within 0.1 au and above 7 $\mearth$. Independently, in a collection of 78 systems containing cold gas giants, we discover 5 systems with detected small planets in our small-planet range. Without a completeness correction, this yields a $6.41\%$ probability of outer giants hosting inner small planets. Our completeness-aware methods yield $42\%$, a factor of $\sim$6.5x greater than this raw value. Conversely, the bottom panel of Figure \ref{fig:clean_contours} shows that the set of small inner planets without detected outer companions are sensitive to those companions with a completeness correction less than a factor of 1.5 . Taken together, these results suggest that small planets are correlated with cold Jupiters, but potentially suppressed by warm Jupiters. The two exceptions in our sample are 55 Cnc, which hosts a very warm giant and an ultra-short period super-Earth, and GL 876, which hosts a 2-day sub-Neptune and a 2:1 resonant pair of super-Jupiters at 30 and 60 days. The fact that the two super-Jupiters are in an orbital resonance suggests that they likely formed farther out and then migrated inward \citep{Yu01}, perhaps explaining how the inner sub-Neptune was able to form despite their presence.

\subsection{Metallicity distributions}

\begin{figure*}[ht!]
\begin{center}
\includegraphics[width = 0.495\textwidth]{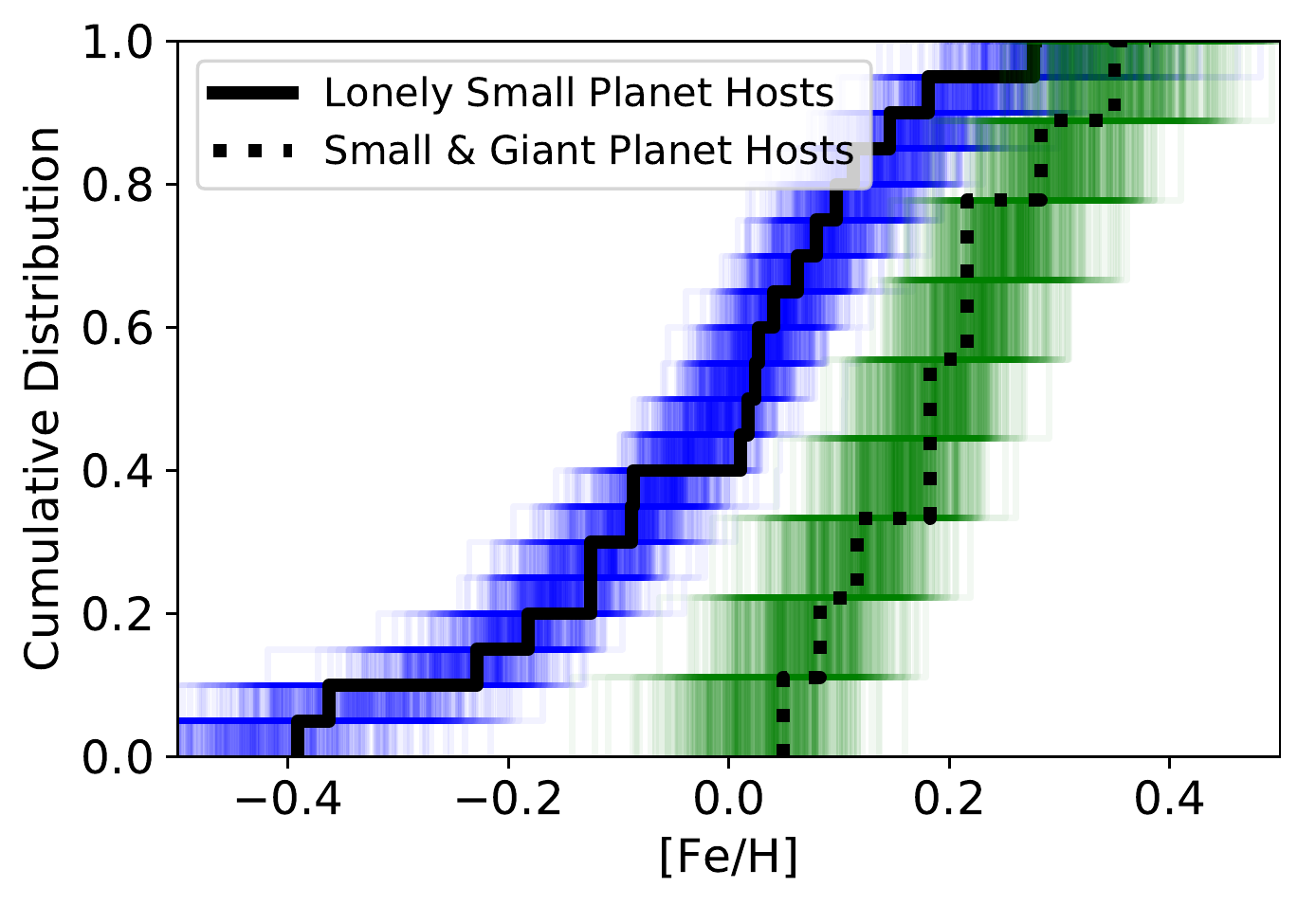}
\includegraphics[width = 0.495\textwidth]{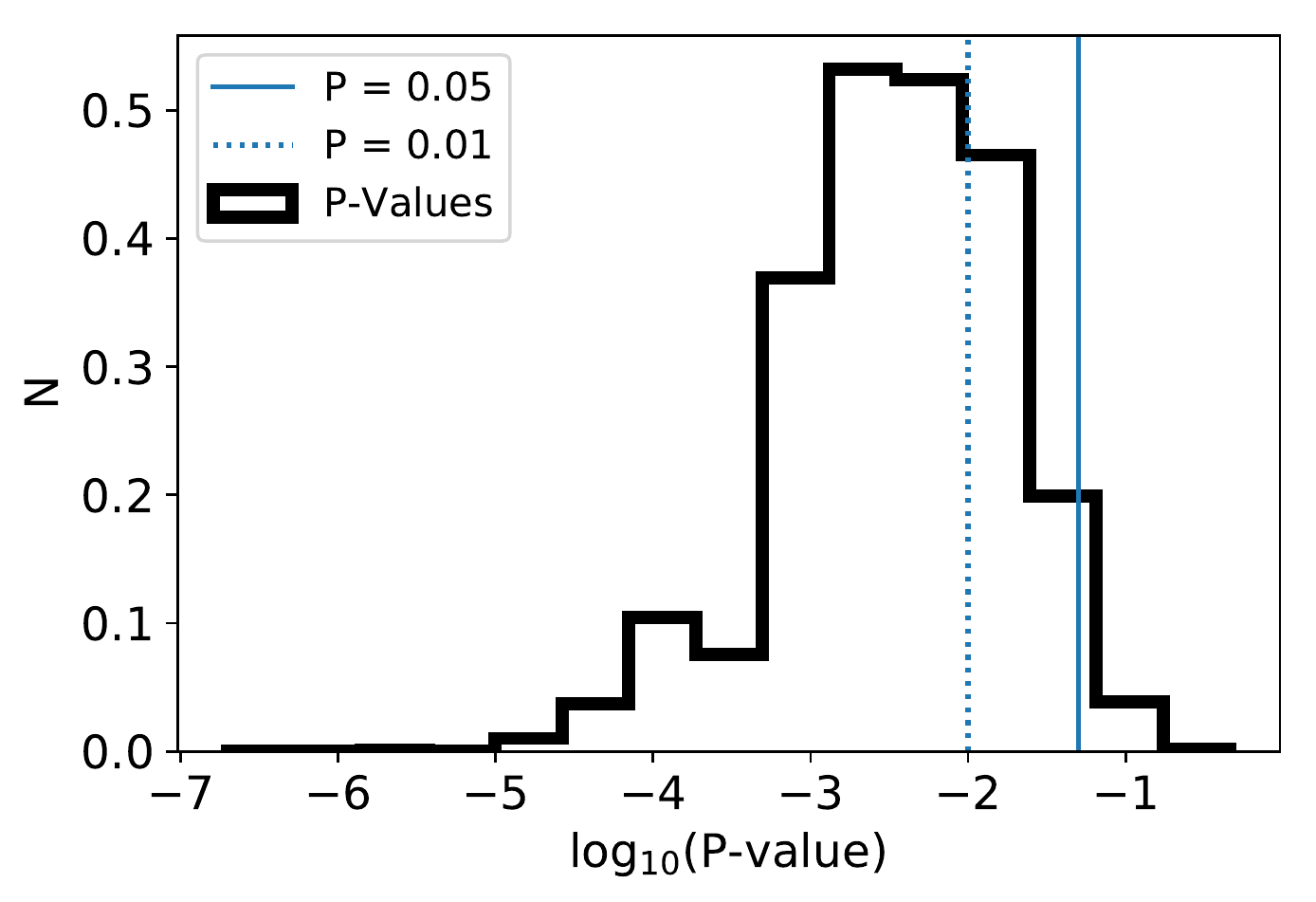}
\caption{Left: Cumulative metallicity distributions for hosts of lonely small planets and hosts of both inner small and outer giant planets. Solid and dashed steps show distributions of median metallicity measurements, while transparent steps show metallicities drawn many times from Gaussian distributions, with means and standard deviations taken from measurement means and uncertainties. Right: A distribution of P-values from many Kolmogorov-Smirnov tests, performed on $5 \times 10^5$ drawn sets of metallicities for the two stellar host groups. 97$\%$ of the draws produce $P < 0.05$, and 73$\%$ of the draws produce $P < 0.01$. This implies that the underlying metallicity distributions of these two groups are distinct.}
\label{fig:metallicity}
\end{center}
\end{figure*}

We used our sample to reproduce the previously derived result \citep{Zhu18, Bryan19} that small planet hosts with outer gas giants are consistently more metal-rich with 97$\%$ significance than hosts of lonely small planets, as seen in Figure \ref{fig:metallicity}. This phenomenon agrees with the well-established correlation between metallicity and giant planet formation and, therefore, may be independent of the presence of small planets. However, since the CLS sample contains few systems with both outer giants and inner small planets, it is difficult to test the reverse effect and determine whether giant planet hosts with small planets have a different metallicity distribution than lonely giant planet hosts.

\section{Discussion} \label{sec:discussion}

\subsection{Reconciling our results with other occurrence work and known systems}

\cite{Bryan19} found that $P(I|J) = 102^{+34}_{-51}\%$, while \cite{Zhu18} found that $P(I|J) = 90\pm 20\%$. Our measurements of $P(I|J) = 32^{+24}_{-16}\%$ and $P(I|O) = 42^{+17}_{-13}\%$ are consistent with the 2019 measurement, but are $\>2\sigma$ inconsistent with the 2018 measurement. Additionally, both of our measurements are more than 2.5$\sigma$ separated from 100$\%$.

Our finding that gas giants within a certain mass and semi-major axis range suppress the formation of inner small planets is highly conditioned on these mass and semi-major axis ranges. We limited this analysis to giant planets within 0.3--3 au. Conversely, \cite{Huang16} found that half of all warm Jupiters have small planet companions by performing a similar analysis on the \textit{Kepler} sample. They defined a warm Jupiter as a giant planet within 10--200 days. 200 days corresponds to a $\sim$0.67 au orbit around a solar mass star, which is only beyond the inner limit of our $'$lukewarm$'$ range by about a factor of 2. This implies that our two results are not necessarily incompatible. Rather, they are drawn from mostly separate giant planet populations, which may have distinct formation or migration mechanisms.

Additionally, while the CLS does not contain 0.3--3 au giant companions to small planets, the \textit{Kepler} sample contains several known systems that fit this description. For instance, the Kepler-167 system contains a 1 $\mjup$ giant at 1.9 au with three super-Earths, and the \textit{Kepler}-1514 system contains a 5 $\mjup$ giant at 0.75 au with an inner 1.1 $\rearth$ planet at 0.1 AU \citep{Kipping16, Dalba21}. Several other \textit{Kepler} systems contain planets that satisfy or almost satisfy our criteria for small-and-giant pairs \citep{Morton16, Holczer16}, as well as non-\textit{Kepler} systems \citep{Bouchy09, Stassun17}. We have not claimed that 0.3--3 au giants completely prevent the formation of small inner planets, only that these giants host inner small planets within 2--30 $\mearth$ with a significantly smaller frequency than giant planets outside this range. Looking to the future, long-baseline RV follow-up of a very large sample of hosts of close-in small planets, such as a subsample of the \textit{TESS} survey \citep{Ricker15}, may uncover a larger number of outer giant companions. This would help clarify the precise distribution of these companions in mass and semi-major axis space.

\subsection{Comparison between direct measurement and Bayesian inference of $P(I|O)$}

Figure \ref{fig:bayes_simple} shows that our direct estimate of $P(I|O)$ is lower than our indirect estimate using Bayes theorem, which calculates $P(I|O)$ as a function of $P(O|I)$, $P(O)$, and $P(I)$. These probabilities are likely mismatched because they assume uniform occurrence across giant planet parameter space, and Figure \ref{fig:checkerboxes} shows that this is not the case. While our broad sample of giant planets fills \msini\ and semi-major axis space, we found no outer companions to small planets among our warm Jupiters, as discussed in Subsection \ref{sec:suppression}. This means that our population of outer giant companions and broader giant planet sample are distinct in parameter space, and that choosing a wide swath of \msini\ and semi-major axis space for our Bayesian inference is not justified. This would also explain why our Jupiter analog comparison, shown in Figure \ref{fig:bayes_jovian}, shows a closer match between a direct measurement and Bayesian inference. 3--7 au and 0.3--13 \mjup\ is a narrower range of parameter space, and well separated from the warmer giants that appear to suppress small planet formation. This difference in giant classification could explain why the two posteriors more closely agree for the narrow definition of Jupiter analogs than for the broader definition that includes all cold gas giants.

\subsection{The nature of cold giant companions}

\begin{figure}[ht!]
\includegraphics[width = 0.5\textwidth]{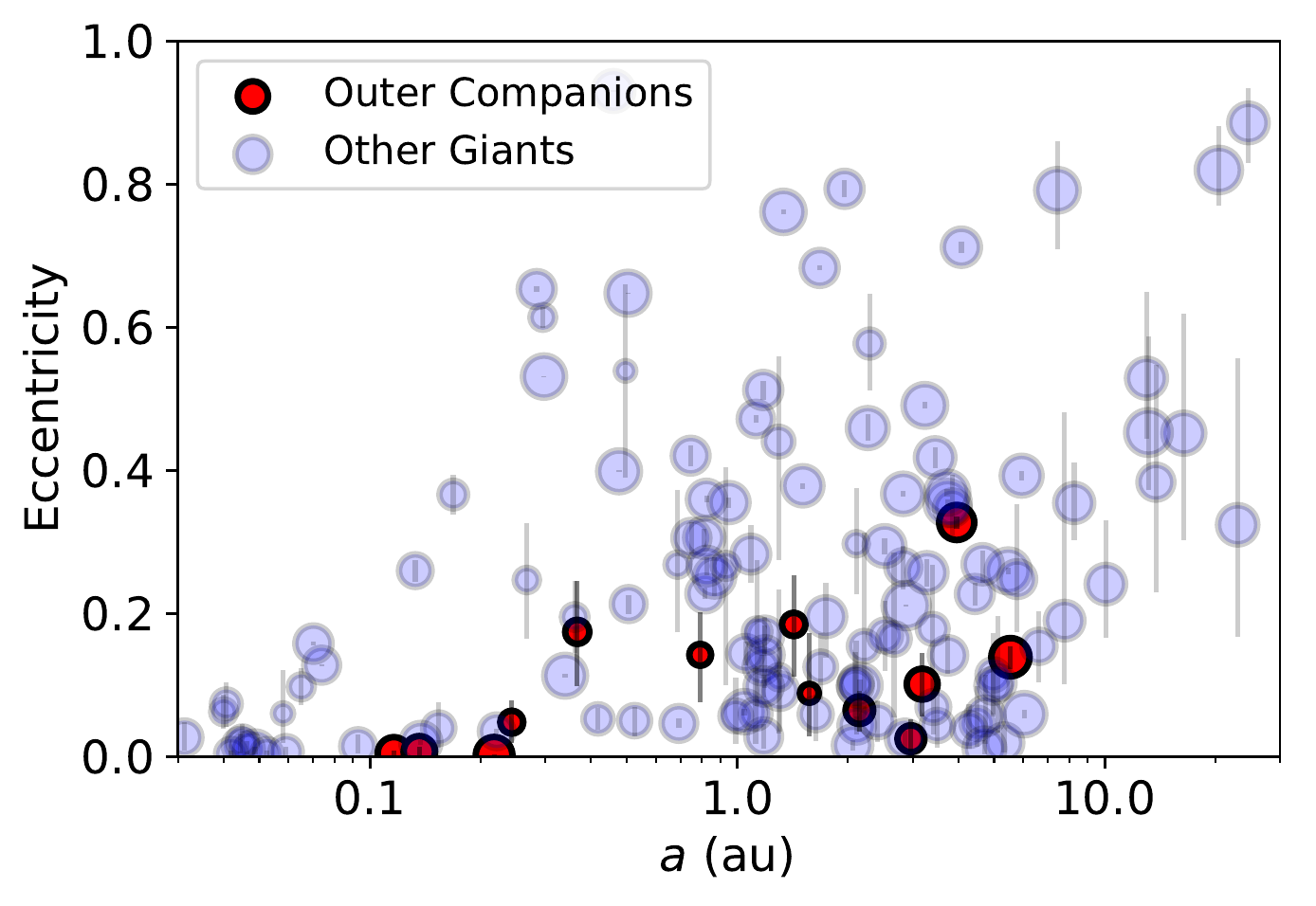}
\caption{A comparison between outer giant companions and the rest of the CLS giant sample in semi-major axis and orbital eccentricity space. Circle size is proportional to log(\msini). Companions to inner small planets are less eccentric than the parent sample, and less massive.}
\label{fig:giant_comparison}
\end{figure}

\begin{figure}[ht!]
\includegraphics[width = 0.5\textwidth]{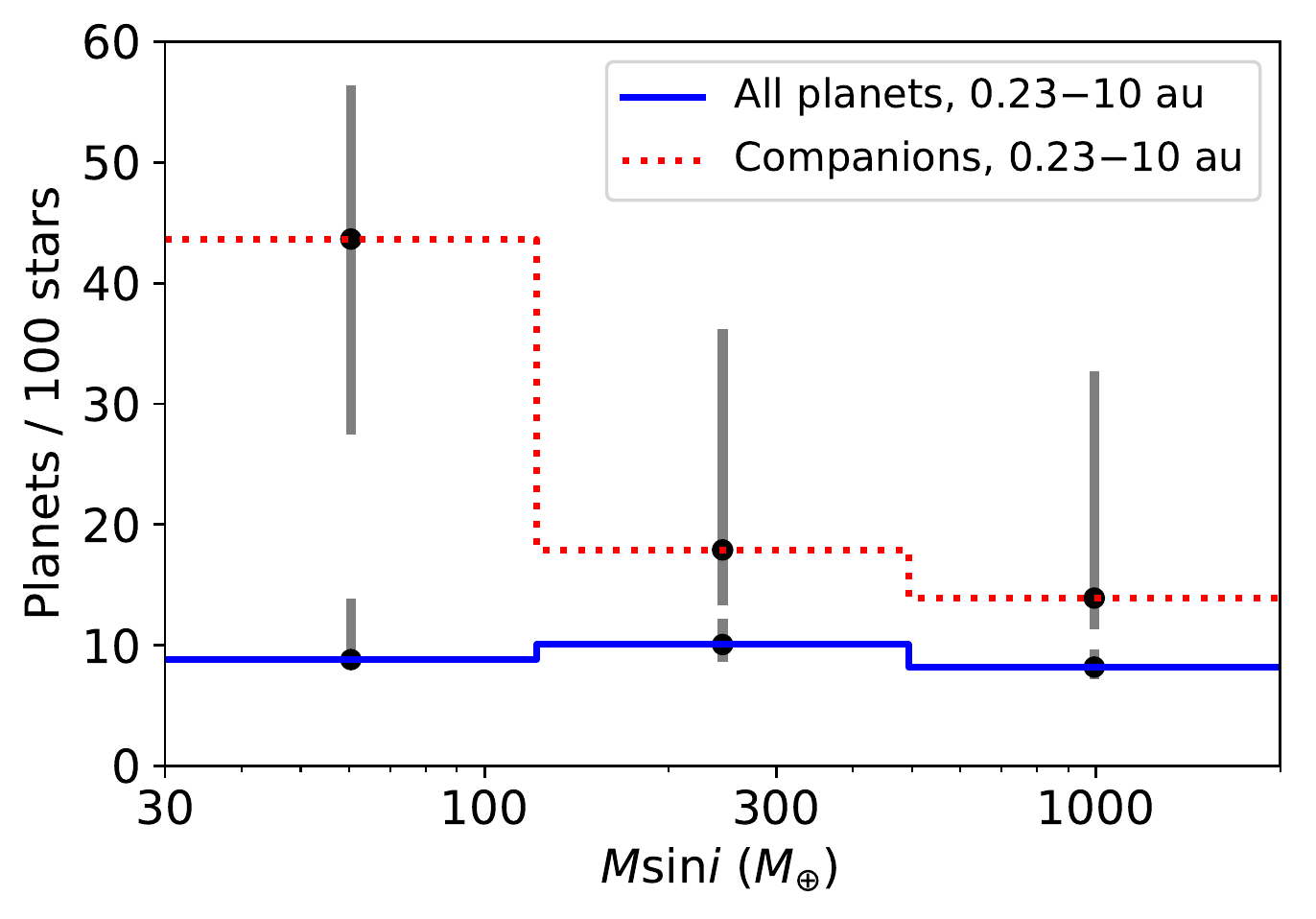}
\caption{Mass functions for all planets and outer companions within 0.23--10 au, produced by marginalizing the occurrence grids in Figure \ref{fig:checkerboxes} along their orbital separation axes. Steps and dots show maximum a posteriori values; vertical bars show 15.9--84.1$\%$ confidence intervals.}
\label{fig:mass_functions}
\end{figure}

\begin{figure}[ht!]
\includegraphics[width = 0.5\textwidth]{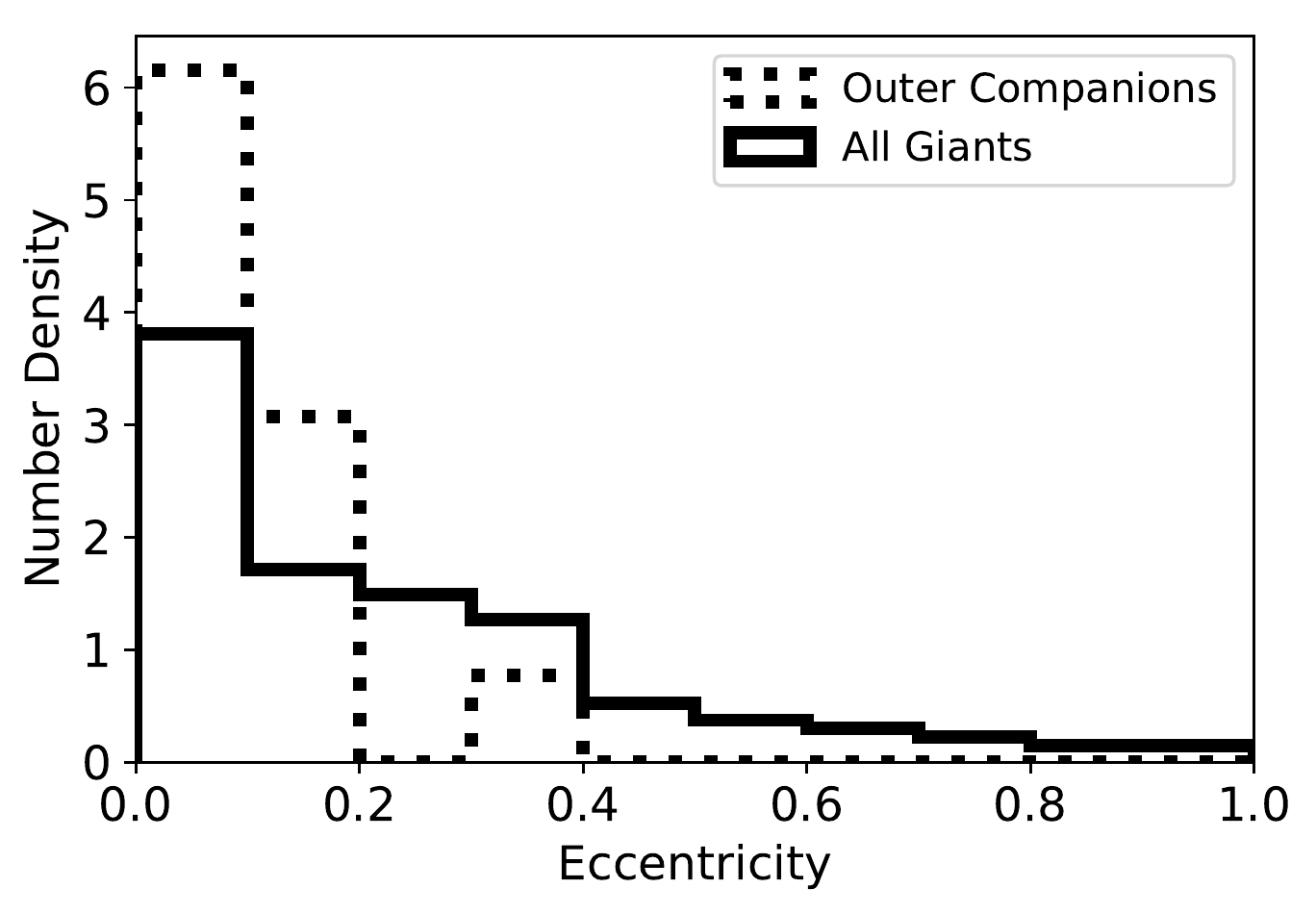}
\caption{Observed eccentricity distributions for all giant planets and outer giant companions. This plot shows histograms of the maximum a posteriori eccentricities of our sample.}
\label{fig:eccentricity_dist}
\end{figure}

Figure \ref{fig:giant_comparison} shows both all giant planets and outer giant companions to small planets in eccentricity, \msini, and semi-major axis space. The outer companions have an upper limit on eccentricity within 0.4, whereas the broader sample follows the beta distribution first described in \cite{Kipping13}. Figure \ref{fig:mass_functions} marginalizes over the occurrence distributions shown in Figure \ref{fig:checkerboxes} to produce mass functions for these two populations within 0.23--10 au. This marginalization shows that outer companions are more frequently found at lower masses than the broader giant sample, with $\sim 2 \sigma$ significance. Figure \ref{fig:eccentricity_dist} shows histograms of the maximum a posteriori eccentricities of all giant planets and outer giant companions to small planets. The broad sample has a moderate-to-high-eccentricity tail that is not shared by the outer companions. This makes intuitive sense, since eccentric giants are disruptive to the inner regions of a planetary system, and can disrupt the early-stage formation of small planets by sweeping through protoplanetary disks, or dynamically scatter small planets.

\subsection{An aside on our sample selection criteria}

Our selection criteria differ from those of \cite{Bryan19} in a number of ways. Our lowest mass planet has \msini\ equal to 3.57 $\mearth$, below which we are almost entirely insensitive even to close-in planets.  We therefore select a mass range of 2--30 $\mearth$, as opposed to 1--10 $\mearth$. We chose our upper limit on \msini\ as an estimate of the mass threshold for runaway gas accretion \citep{Lissauer09}. This limit also happens to correspond to a possible valley in the mass-radius distribution of planets, as seen in \cite{Neil2020}. In order to test whether this difference will significantly challenge our comparison to the prior results, we recompute our measurements with tighter limits on small planet parameter space, moving to 2--20 $\mearth$ and 0.023--0.5 au. We show these results in Table \ref{tab:conditionals_full}. $P(I|O)$ is consistent for both definitions of an inner small planet; $P(O|I)$ is $\sim 1 \sigma$ distinct. We cannot compare Jupiter analog conditional probabilities because the narrower sample of small planets does not include any companions to Jupiter analogs. These results imply that our choice of \msini\ and $a$ limits for small inner planets do not significantly impact our comparisons to studies with different definitions of small planets. We also explored the impact of changing our lower \msini\ limit from 2 $\mearth$ to 3 $\mearth$, since there are no companion small planets in our sample that are less massive than 3 $\mearth$. These changes decreased $P(I|O)$ and $P(I)$ by less than $1 \sigma$ and $1.5 \sigma$ respectively, as seen in Table \ref{tab:conditionals_full}.

\begin{deluxetable*}{lrrrr}
\caption{Same as Table \ref{tab:conditionals}, but middle column is for small planets within 2--20 $\mearth$ and 0.023--0.5 au, while right-hand columns are for 3--30 $\mearth$ and 6--30 $\mearth$. {\label{tab:conditionals_full}}}
\tabletypesize{\large}
\tablehead{\colhead{Condition} & (1 au, 2--30 $\mearth$) & (0.5 au, 2--20 $\mearth$) & (0.5 au, 3--30 $\mearth$) & (0.5 au, 6--30 $\mearth$)}
\startdata
Inner & $0.276^{+0.058}_{-0.048}$ & $0.281^{+0.066}_{-0.051}$ & $0.191^{+0.036}_{-0.035}$ & $0.143^{+0.030}_{-0.026}$ \\
Outer$|$Inner & $0.41^{+0.15}_{-0.13}$ & $0.29^{+0.14}_{-0.11}$ & $0.43^{+0.17}_{-0.13}$ & $0.42^{+0.15}_{-0.12}$ \\
Inner$|$Outer & $0.42^{+0.17}_{-0.13}$ & $0.46^{+0.20}_{-0.16}$ & $0.28^{+0.12}_{-0.09}$ & $0.219^{+0.080}_{-0.075}$ \\
Jupiter$|$Inner & $0.133^{+0.097}_{-0.063}$ & No detections & $0.21^{+0.12}_{-0.09}$ & $0.21^{+0.11}_{-0.09}$ \\
Inner$|$Jupiter & $0.32^{+0.24}_{-0.16}$ & No detections & $0.25^{+0.17}_{-0.13}$ & $0.16^{+0.13}_{-0.07}$ \\
Outer & $0.176^{+0.024}_{-0.019}$ \\
Jupiter & $0.072^{+0.014}_{-0.013}$ \\
\enddata
\end{deluxetable*}

Additionally, we did not search for planets with orbital periods less than one day, since this would produce alias issues in our automated search pipeline. This leads to a complication regarding 55 Cnc, which hosts a super-Earth with an orbital period of 0.74 days. This planet is the only previously known USP in our sample, and this system is one of the few that we initialized with known planets in our search, including a Keplerian orbit for the USP in order to properly model our RV data. This system also stands out from the rest of our sample in a number of other ways, such as hosting both a hot Jupiter and multiple outer, less massive giants. Since our blind search does not extend below 1 day, we should in principle limit our small-planet sample to planets beyond 0.02 au, which corresponds to just over a 1-day orbit around a G dwarf. This excludes 55 Cnc and its giant planets from our Bayesian estimates of inner and outer companion probability and leaves only 2 planets within the Jupiter analog box instead of 3. We opted to exclude 55 Cnc from our conditional probability analysis for the sake of consistency, but left it in our outer giant occurrence grids shown in Figure \ref{fig:checkerboxes}. Redoing the analysis with an inner limit of 0.015 au, so that 55 Cnc is included, we find that the probability of hosting a close-in small planet given the presence of an outer Jupiter analog is $0.39^{+0.21}_{-0.16}$, as opposed to $0.32^{+0.24}_{-0.16}$ without 55 Cnc. Likewise, when including 55 Cnc, the probability of hosting a close-in small planet given the presence of a cold gas giant in broader parameter space (0.23--10 au, 30--6000 $\mearth$) is $0.42^{+0.17}_{-0.13}$, as opposed to $0.42^{+0.17}_{-0.12}$ without 55 Cnc, i.e., nearly identical.

Alongside 55 Cnc, GJ 876 is the one other system in our sample that hosts both a detected small planet and warm gas giants. This system hosts a small planet on a 2-day orbit and two giant planets in a 2:1 resonance at 30 and 60 days. We propose that this system is the exception that supports our theory of warm Jupiters suppressing inner small planet formation, since this resonant pair may have migrated inward from beyond 1 au \citep{Yu01, Batygin15, Nelson16}.

\subsection{An aside on multiplicity bias} \label{subsec:bias}

We investigate whether our method for estimating the probability that a star hosts at least one planet ($P(1+)$) in a given parameter space is systematically biased in cases of high planet multiplicity. We estimate the possible magnitude of this effect using a Monte Carlo experiment to recover the true value of $P(1+)$ given a toy model for planet multiplicity and simulated observed population. First, we choose an underlying probability of hosting at least one planet $p$, as well as a simple multiplicity distribution $f_n$ given the presence of at least one planet, capped at 3 planets.

Then, we perform a single step of our Monte Carlo experiment by `creating' 81 stars, the size of our outer giant $O$ host sample. Each star has a probability $p$ of hosting any inner small planets, and probability $f_n$ of hosting $n$ such planets, if it hosts any such planets at all. We sample these planets from uniform ln($a$) and ln($M$) distributions in the desired parameter space, in our case our small planet definition. We then determine how many planets we detect around each star by only keeping the planets with ln($a$) and ln($M$) pairs that have search completeness higher than a random number drawn from $U(0, 1)$. We randomly select a new completeness contour from the outer giant host sample for each generated system.

\begin{figure}[ht!]
\includegraphics[width = 0.5\textwidth]{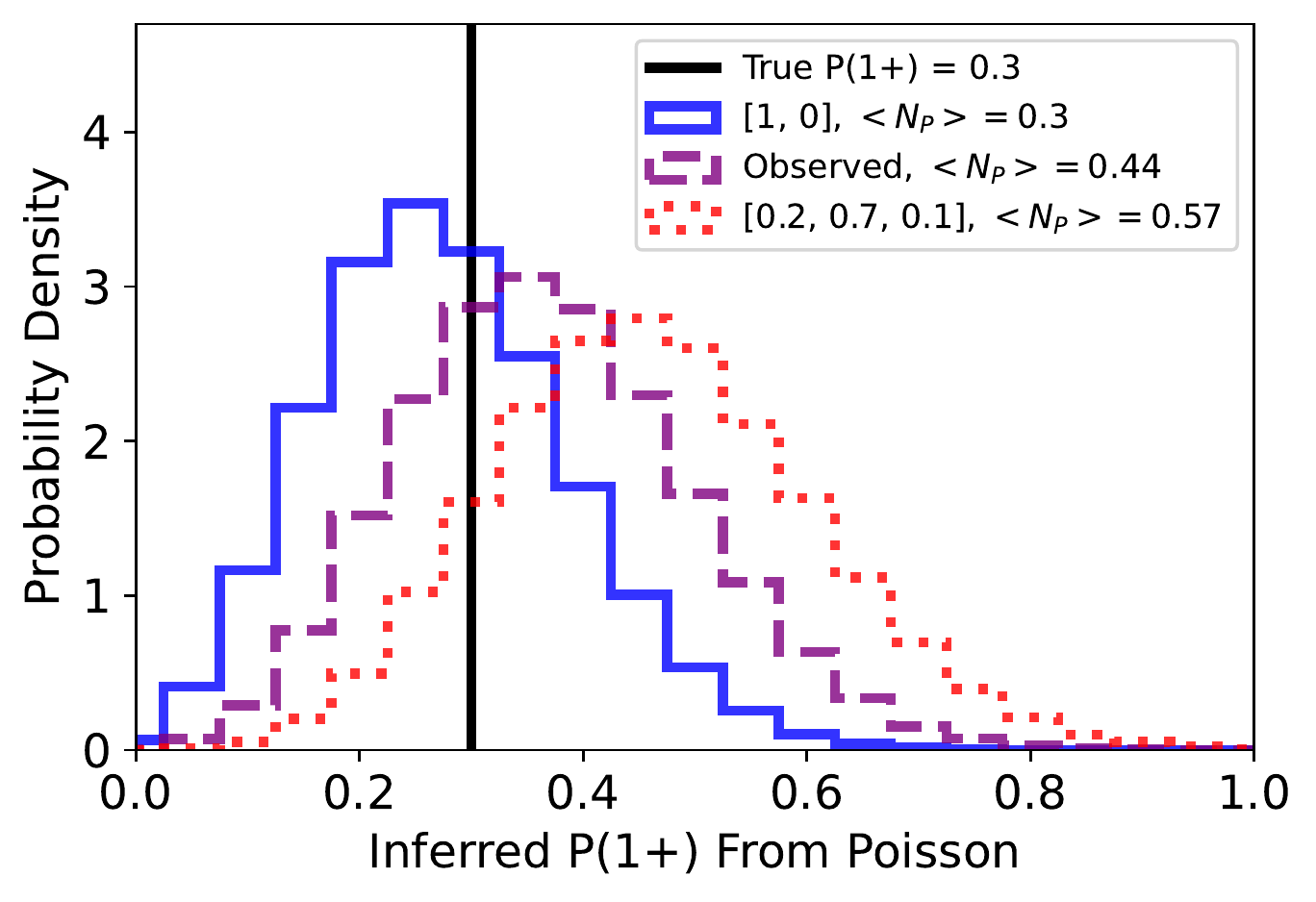}
\caption{The results of three Monte Carlo experiments detailed in Subsection \ref{subsec:bias}. We estimate how planet multiplicity biases our estimates of the probability that a star hosts at least one small planet, or $P(1+)$. The black vertical line is the true value of $P(1+)$ for all three experiments. The legend shows the multiplicity distribution and resulting average number of planets per host $<N_P>$ for all three experiments. In the one-per-host case, the Poisson model accurately recovers $P(1+)$, while cases with higher $<N_P>$ increasingly overestimate the true value.}
\label{fig:poisson_monte}
\end{figure}

Once we have our population of observed planets, we run our Poisson likelihood model on the first-observed planets to generate our estimate of the probability that a star hosts at least one planet in a given parameter space, or $P(1+)$. Figure \ref{fig:poisson_monte} shows our results for $p=0.3$ and three different choices of multiplicity, including the distribution of detected multis shown in Figure \ref{fig:multiplicity}. When we assume one small planet per host, our Poisson model accurately retrieves the underlying probability of hosting at least one planet. With increasing average multiplicity, our model marginally overestimates $P(1+)$ by a small but increasing factor. This is because increasing the expected intrinsic number of planets per host increases the probability of detecting at least one planet around a true host.

Correcting for this bias would require confident knowledge of the multiplicity distribution of small planets, which is currently contested even after much work with \textit{Kepler} \citep[e.g.,][]{Zhu18, He19}. Resolving this issue would require a much larger and more complete small planet sample than the one available through the CLS. Additionally, this bias analysis is specific to our Poisson likelihood methods, which are better suited to correcting for survey incompleteness than binomial or Bernouilli estimates. Hopefully, future precise RV exoplanet surveys will produce more rigorous small planet multiplicity measurements, and create opportunities to better understand bias in occurrence rates due to multiplicity.

\subsection{Implications for planet formation}

One key takeaway from our analysis is that lukewarm Jupiters may either suppress the formation or migration of small inner planets or destabilize the orbits of inner super-Earths. The first conjecture fits with theoretical work \citep{Kley12, Moriarty15} that shows how gas giants that are sufficiently massive or close to their host stars can create gaps in the protoplanetary disk and prevent the inward flow of solids beyond their orbits. This cuts off the supply of material during the critical timescales of pebble accretion, thus depriving rocky cores of the fuel needed to grow into super-Earths or larger planets \citep{Chachan22}. Alternatively, warm or lukewarm gas giants may excite the eccentricities of nascent inner small planets and destabilize their orbits into ejection or accretion \citep{Schlecker20}. Both of these explanations imply that a cold gas giant beyond 5 au such as Jupiter is not detrimental to interior small planet formation, but a Jupiter-mass giant within 0.3--3 au may be.

These ideas do not have to clash with the known coexistence of small planets and warm gas giants. \cite{Huang16} found that warm Jupiters (10--200 days) with close companions are substantially more common than hot Jupiters ($<$ 10 days) with close companions. It is possible that there is a warm Jupiter pile-up due to migration, which takes cold gas giants all the way through the region where we see a dearth of warmer giant companions to small inner planets.

\section{Conclusions and Future Work} \label{sec:conclusions}

We explored the relationship between small, close-in planets and outer giants by computing absolute and conditional probabilities for these two populations. We found that $42^{+17}_{-13}\%$ of stars that host a giant planet within 0.23--10 au also host an inner small planet between 2 and 30 $\mearth$, and that $32^{+24}_{-16}\%$ of stars that host a Jupiter analog (3--7 au, 0.3--13 \mjup) also host an inner small planet between 2 and 30 $\mearth$. These probabilities are $\sim 1 \sigma$ separated from the absolute probability of hosting a small close-in planet, implying an inconclusive effect of outer gas giants on the occurrence of small, close-in companions. On the other hand, the probability of hosting an outer gas giant given the presence of a small planet is $1.65\sigma$ enhanced over the absolute probability of hosting an outer gas giant. We also confirmed the known result that stars with both small, close-in planets and cold giants tend to be more metal-rich than stars with only small planets. Additionally, we used Monte Carlo simulations to estimate how small planet multiplicity might bias Poisson estimates of the probability that a star hosts at least one small planet. We found that multiplicity may result in overestimating this probability, but that assumptions in our Poisson model may reduce the magnitude of this bias.

The next paper in the California Legacy Survey will split our sample into single-giant and multiple-giant systems and investigate the differences and commonalities between these two groups. Taken together with this study of small planets and cold giants, as well as the broader study of gas giants in \cite{Fulton21}, this work may reveal new insights into the formation, evolution, and final architectures of planetary systems.

\acknowledgments
L.J.R.\ led the construction of this paper, including performing all analysis, generating all of the figures, and writing this manuscript. Y.C., F.D., H.A.K., and A.W.H. advised substantially on the scientific direction of this work. All other coauthors contributed feedback on this manuscript and analysis therein.

We thank the anonymous reviewer for constructive feedback on our analysis, particularly regarding the details and pitfalls of measuring planet occurrence. We thank Ken and Gloria Levy, who supported the construction of the Levy Spectrometer on the Automated Planet Finder, which was used heavily for this research. We thank the University of California and Google for supporting Lick Observatory, and the UCO staff as well as UCO director Claire Max for their dedicated work scheduling and operating the telescopes of Lick Observatory.

A.C. acknowledges support from the National Science Foundation through the Graduate Research Fellowship Program (DGE 1842402). G.W.H.\ acknowledges long-term support from NASA, NSF, Tennessee State University, and the State of Tennessee through its Centers of Excellence program. A.W.H.\ acknowledges NSF grant 1753582. H.A.K. acknowledges NSF grant 1555095. P.D.\ gratefully acknowledges support from a National Science Foundation (NSF) Astronomy \& Astrophysics Postdoctoral Fellowship under award AST-1903811.

This work has made use of data from the European Space Agency (ESA) mission {\it Gaia} (\url{https://www.cosmos.esa.int/gaia}), processed by the {\it Gaia} Data Processing and Analysis Consortium (DPAC, \url{https://www.cosmos.esa.int/web/gaia/dpac/consortium}). Funding for the DPAC has been provided by national institutions, in particular the institutions participating in the {\it Gaia} Multilateral Agreement.

\vspace{5mm}
\facilities{HIRES, Lick-Hamilton, APF}

\software{All code used in this paper is available at  \url{github.com/California-Planet-Search/rvsearch} and \url{github.com/leerosenthalj/CLSIII}. This research makes use of GNU Parallel \citep{Tange11}. We made use of the following publicly available Python modules: \texttt{astropy} \citep{Astropy-Collaboration13}, \texttt{matplotlib} \citep{Hunter07}, \texttt{numpy/scipy} \citep{numpy/scipy}, \texttt{pandas} \citep{pandas}, \texttt{emcee} \citep{DFM13}, and \texttt{RadVel} \citep{Fulton18}.}

\bibliographystyle{aasjournal}
\bibliography{CLSIII}{}

\end{document}